\documentclass[10pt,aps,pra,twocolumn,showpacs,amsmath,nofootinbib]{revtex4-1}
\usepackage{bm}
\usepackage{hyperref}
\usepackage{graphicx} 
\usepackage[english]{babel}
\usepackage[T1]{fontenc}
\usepackage{xcolor}
\usepackage{amssymb}

\begin{document}

\title{The RbSr \texorpdfstring{$^2\Sigma^+$}{a} ground state investigated via spectroscopy of hot \& ultracold molecules}

\author{Alessio Ciamei}
\email[Corresponding author: ]{a.ciamei@uva.nl}
\affiliation{Van  der  Waals-Zeeman  Institute,  Institute  of  Physics,  University  of Amsterdam, Science  Park  904,  1098  XH  Amsterdam,  The  Netherlands}
\author{Alex Bayerle}
\email[Corresponding author: ]{A.Bayerle@arcnl.nl}
\altaffiliation[present address: ]{ARCNL, Science Park 110, 1098 XG Amsterdam (NL)}
\affiliation{Van  der  Waals-Zeeman  Institute,  Institute  of  Physics,  University  of Amsterdam, Science  Park  904,  1098  XH  Amsterdam,  The  Netherlands}
\author{Vincent Barb\'e}
\author{Lukas Reichs\"{o}llner}
\author{Chun-Chia Chen}
\author{Benjamin Pasquiou}
\author{Florian Schreck}
\affiliation{Van  der  Waals-Zeeman  Institute,  Institute  of  Physics,  University  of Amsterdam, Science  Park  904,  1098  XH  Amsterdam,  The  Netherlands}
\author{Jacek Szczepkowski}
\email[Corresponding author: ]{jszczep@ifpan.edu.pl}
\author{Anna Grochola}
\author{Wlodzimierz Jastrzebski}
\affiliation{Institute of Physics, Polish Academy of Sciences, Al. Lotnik\'ow 32/46, Warszawa 02\textendash668, Poland}
\author{Slava M. Tzanova}
\affiliation{Institut f\"ur Quantenoptik und Quanteninformation (IQOQI), \"Osterreichische Akademie der Wissenschaften, 6020 Innsbruck, Austria}
\author{Pawel Kowalczyk}
\affiliation{Institute of Experimental Physics, Faculty of Physics, University of Warsaw, ul.~Pasteura~5, 02-093 Warszawa, Poland}

\begin{abstract}
We report on spectroscopic studies of hot and ultracold RbSr molecules, and combine the results in an analysis that allows us to fit a potential energy curve (PEC) for the $\mathrm{X}(1)^2\Sigma^+$ ground state bridging the short-to-long-range domains. The ultracold RbSr molecules are created in a $\mu\mathrm{K}$ sample of Rb and Sr atoms and probed by two-colour photoassociation spectroscopy. The data yield the long-range dispersion coefficients $C_6$ and $C_8$, along with the total number of supported bound levels. The hot RbSr molecules are created in a $1000 \,\mathrm{K}$ gas mixture of Rb and Sr in a heat-pipe oven and probed by thermoluminescence and laser-induced fluorescence spectroscopy. We compare the hot molecule data with spectra we simulated using previously published PECs determined by three different \textit{ab-initio} theoretical methods. We identify several band heads corresponding to radiative decay from the $\mathrm{B}(2)^2\Sigma^+$ state to the deepest bound levels of $\mathrm{X}(1)^2\Sigma^+$. We determine a mass-scaled high-precision model for $\mathrm{X}(1)^2\Sigma^+$ by fitting all data using a single fit procedure. The corresponding PEC is consistent with all data, thus spanning short-to-long internuclear distances and bridging an energy gap of about $75\%$ of the potential well depth, still uncharted by any experiment. We benchmark previous \textit{ab-initio} PECs against our results, and give the PEC fit parameters for both $\mathrm{X}(1)^2\Sigma^+$ and $\mathrm{B}(2)^2\Sigma^+$ states. As first outcomes of our analysis, we calculate the \textit{s}-wave scattering properties for all stable isotopic combinations and corroborate the locations of Fano-Feshbach resonances between alkali Rb and closed-shell Sr atoms recently observed \href{https://www.nature.com/articles/s41567-018-0169-x}{[Barb\'e \textit{et al.}, Nat. Phys., 2018, DOI:10.1038/s41567-018-0169-x]}. These results and more generally our strategy should greatly contribute to the generation of ultracold alkali -- alkaline-earth dimers, whose applications range from quantum simulation to state-controlled quantum chemistry.
\end{abstract}

\maketitle

\section{Introduction}
\label{sec:Intro}
Production of ultracold molecules composed of one alkali and one alkaline-earth(-like) atom is being pursued with increasing effort over the last years, boosted by the achievement of quantum degeneracy for gases of alkaline-earth atoms and atoms with similar electronic structure \cite{Stellmer2009bec, Taie2012, Kraft2009CaFirstBEC}. These heteronuclear open-shell molecules possess a $^2\Sigma$ electronic ground state. In the rovibronic ground state, they exhibit a non-zero electronic spin angular momentum and a strong permanent electric dipole moment. These properties make them suitable for quantum simulations of magnetism and topological quantum phases mediated by the induced electric dipole-dipole interaction \cite{Perez-Rios2010MagneticExciton, Micheli2006ToolboxPolMol, Baranov2012RevDipQGases,Bohn1002RevColdMol,KremsBookColdMolecules}. Molecules with $^2\Sigma$ ground state could also be used as sensitive magnetic field sensors \cite{Alyabyshev2012BfieldImagingPolarMolecules}, quantum computing platforms \cite{Herrera2014QuantumGate2Sigma}, and probes of parity-violations and variation of the proton-to-electron mass ratio \cite{Cahn2014ParityViolation, Kajita2008ProtonTonElectronMass2Sigma, Safronova2017RevNewPhysicsAtMol}. If one can produce a quantum degenerate gas of molecules, where all degrees of freedom are under control, one can study quantum chemical reactions and their dynamics at the most fundamental level, with full control over the reactants, in dependence of electromagnetic fields, and detecting reaction products \cite{Wolf2017StateToStateChemistry, Ospelkaus2010ChemReactKRb, Sikorsky2018SpinAtomIonChem, Krems2008RevColdChemistry,Bohn1002RevColdMol,KremsBookColdMolecules}.

In order to create molecules at ultracold temperatures and to understand quantum chemistry processes, an accurate molecular model is needed. Recently \textit{ab-initio} calculations for alkali -- alkaline-earth(-like) molecules have provided potential energy curves (PECs), permanent electric dipole moments and transition dipole moments, and a few attempts at benchmarking theories with experiments have been recorded \cite{JacekKSR, Pototschnig2015RbCa, Ivanova2011LiCHeatpipe, Schwanke2017LiSrHeatpipe, Borkowski2013ScatLengthRbYb, Krois2013LiCaHelium, Pichler1989LiMg}. The precision of \textit{ab-initio} calculations is typically not enough to reliably predict the properties that need to be known to form ultracold molecules, such as molecular binding energies. Theory must therefore be complemented by spectroscopy experiments.

Different spectral ranges can be explored with the help of various types of spectroscopy, three of which being relevant for the present work. Photoassociation (PA) spectroscopy of ultracold atoms provides data with precision and accuracy reaching down to the $\mathrm{kHz}$ level \cite{Jones2006RevPA, Borkowski2016GravityPA, Borkowski2017ConfGravityPA}. PA spectroscopy favours the production of weakly-bound molecules, since their wavefunction has the best overlap with the large wavefunction describing colliding atoms. Knowledge of these weakly-bound levels is sufficient to determine the long-range behaviour of the PECs \cite{Muenchow2011TwoColorPA, Borkowski2014MassScalingSr2}. Thermoluminescence and laser induced fluorescence (LIF) spectroscopy in high-temperature ovens provide spectra with many optical lines at a fraction of $\mathrm{cm}^{-1}$ precision \cite{Tiemann2017KCa, Tiemann2017LiSr, JacekKSR}. Thermoluminescence and LIF spectra are usually dominated by the radiative decay towards the most bound levels of the ground-state potential and therefore allow to determine the behaviour of the PECs in a range of internuclear distances centred around the potential equilibrium distance.

In this paper we present two independent experimental investigations of alkali -- alkaline-earth RbSr molecules, two-colour PA spectroscopy of ultracold $\mathrm{Rb}\operatorname{-}\mathrm{Sr}$ mixtures, and thermoluminescence/LIF spectroscopy of hot molecules, both carried out for the first time on this system. By combining the results from both experiments in a joint analysis and exploiting three previously reported state-of-the-art \textit{ab-initio} calculations \cite{PiotrOlivierRbSr, ErnstLatestPECs}, we can provide a PEC for RbSr ground-state molecules representing accurately all our experimental data and smoothly bridging the gap between the two spectral ranges investigated. We also determine the molecular constants of the $\mathrm{X}(1)^2\Sigma^+$ and $\mathrm{B}(2)^2\Sigma^+$ states, and dispersion coefficients of the $\mathrm{X}(1)^2\Sigma^+$ state. We use these fitted PECs to benchmark the \textit{ab-initio} calculations, which come from three independent theoretical methods. Thanks to this analysis, we can infer the molecular spectra with sufficient accuracy to guide future experiments (e.g. STIRAP path towards rovibronic ground state \cite{Debatin2011SpectroRbCs, WangRbSr, Guo2017NaRbSpectro}), as well as atomic properties such as scattering cross sections and magnetic Fano-Feshbach resonances.

This manuscript is structured as follows. In section \ref{sec:DescriptionRbSrSpectrum}, we summarize the experimental and theoretical information currently available on molecular RbSr and we introduce the molecular potentials that we investigate. In section \ref{sec:ColdExpPart}, we present two-colour PA spectroscopy of three RbSr isotopologues performed on $\mu\mathrm{K}$ atomic mixtures. We use the PA spectroscopy results to fit a model from which we extract the long-range dispersion coefficients $C_6$ and $C_8$ along with the zero-energy semi-classical action. Based on this spectroscopy type alone, we calculate the \textit{s}-wave scattering properties of all isotopic combinations of Rb and Sr, and explain the location of magnetic Fano-Feshbach resonances observed in previous work by some of the authors \cite{RbSrFFR}. The Fano-Feshbach resonances are then included in the fit to provide a single comprehensive model. We corroborate this analysis by comparison with independent cross-thermalization experiments. In section \ref{sec:HotExpPart}, we present the thermoluminescence spectroscopy and LIF spectroscopy in a $1000\,\mathrm{K}$ heat-pipe oven. We detail the production of the molecular gas sample and its interrogation. We compare the recorded spectrum with three simulated spectra recreated starting from three \textit{ab-initio} theory calculations. From this comparison we identify 24 band heads in the data and give the fitted Dunham coefficients, which describe the lowest vibrational energy levels of the ground and first excited $^2\Sigma^{+}$ states. In section \ref{sec:JointPart}, we use the results from both types of spectroscopy to refine the three \textit{ab-initio} ground-state potentials via a direct potential fit of an analytic function. We discuss the final results and how they compare with theory. In section \ref{sec:Conclusions}, we conclude and give an outlook.

\section{\texorpdfstring{R\MakeLowercase{b}S\MakeLowercase{r}}{RbSr} state of the art}
\label{sec:DescriptionRbSrSpectrum}

We first introduce the molecular structure of RbSr and the results of previous studies. RbSr has recently been the subject of theoretical works \cite{Guerout2010AlkSrPECandPDM, PiotrOlivierRbSr, ErnstLatestPECs, KajitaRbSr, WangRbSr}, two of which \cite{PiotrOlivierRbSr, ErnstLatestPECs} cover the spectral region that we are investigating and provide state-of-the-art PECs based on \textit{ab-initio} calculations. In ref.\,\citenum{PiotrOlivierRbSr}, \citeauthor{PiotrOlivierRbSr} compare two different methods. The first is a full-configuration-interaction (FCI) treatment of RbSr, represented as a molecule with 3 valence electrons subject to an effective core potential (ECP) complemented with a core polarization potential (CPP), which is referred to as FCI-ECP+CPP. The second is a spin-restricted coupled-cluster (RCC) method, applied to a 19 electron problem subject to a fully-relativistic small-core ECP with single, double and triple excitations, referred to as RCCSD(T). In ref.~\citenum{ErnstLatestPECs}, \citeauthor{ErnstLatestPECs} provide PECs obtained via multiconfigurational self-consistent field calculations, involving ECP and CPP, followed by second order multireference configuration interaction, which we label MRCI in the following. For all three methods, PECs of the non-rotating molecule are calculated without or with inclusion of the fine-structure Hamiltonian, resulting in Hund case (a,b)\footnote[1]{Hund cases (a) and (b) are equivalent for non-rotating molecules.} or (c) representation, respectively.

Experimental investigation of RbSr has been restricted so far to Helium-nanodroplet-assisted spectroscopy \cite{ErnstHeNPRL, ErnstPCCP}. In these experiments a supersonic jet of He droplets is sequentially injected into pickup cells containing Rb or Sr, which can get caught on the droplet surface and reactively collide forming a RbSr molecule. In contact with superfluid He, RbSr further relaxes to its vibronic ground state, which greatly simplifies spectroscopic studies. Extensive spectroscopy data were collected via resonance-enhanced two-photon ionization, elucidating the electronic structure of RbSr in the spectral region $11600-23000\,\mathrm{cm}^{-1}$, where the precision was limited by line-broadening due to the coupling of RbSr to the He droplet. Remarkably, RbSr desorbed from the droplet upon laser excitation, allowing to record fluorescence of free RbSr molecules and to extract the harmonic constant of the ground state. The experimental value was consistent with theoretical predictions, however the measurement precision was not sufficient to discriminate between the three aforementioned high-precision theoretical PECs.

\begin{figure}[ht]
\centering
  \includegraphics[width=\columnwidth]{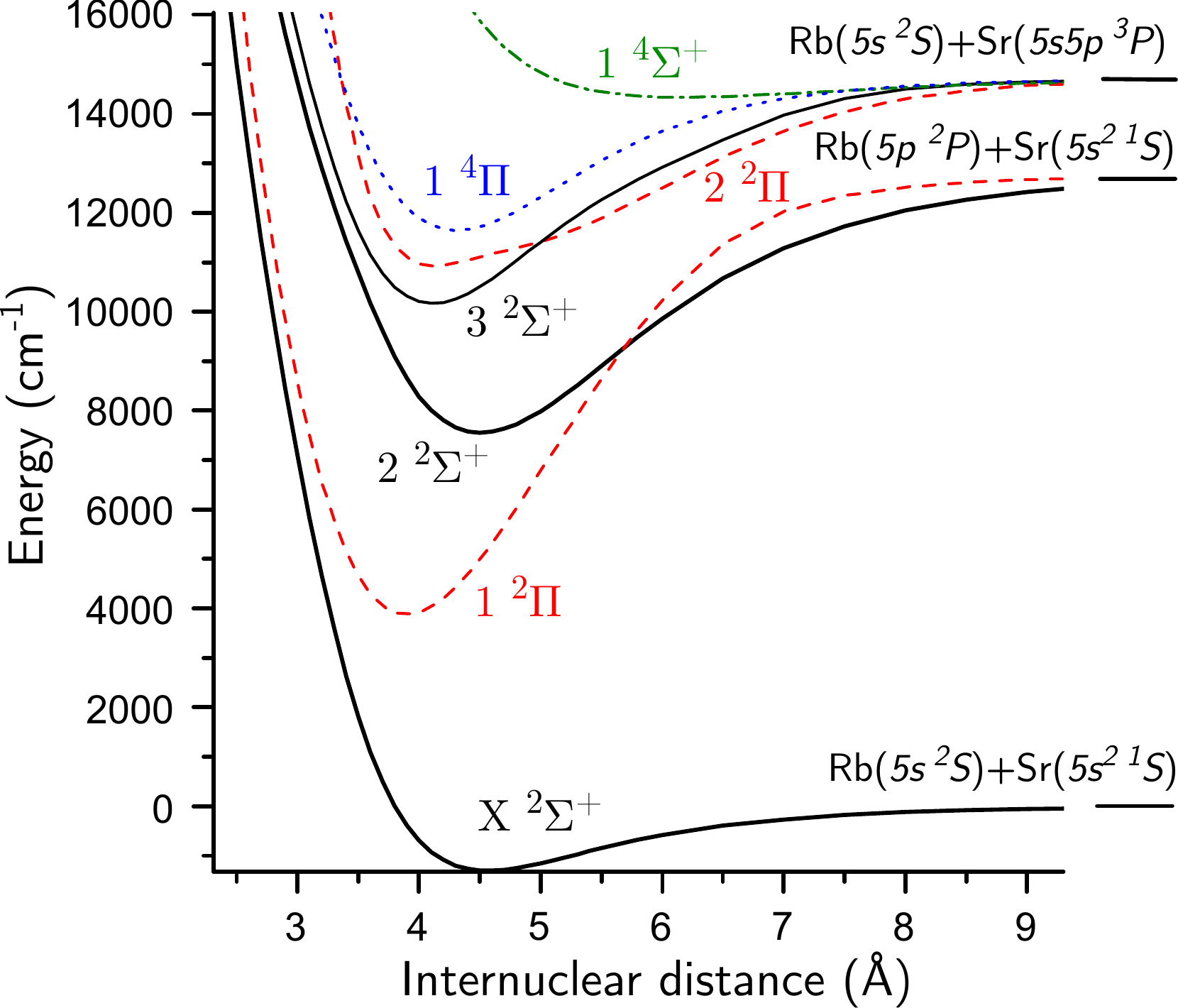}
  \caption{The potential energy curves calculated using the MRCI method \cite{Ernst2014JCP} for all states correlating to the $\mathrm{Rb} (5s\;^2S)+\mathrm{Sr} (5s^2\;^1S)$, $\mathrm{Rb} (5p \; ^2P) + \mathrm{Sr} (5s^2 \; ^1S)$ and $\mathrm{Rb}(5s \; ^2S) + \mathrm{Sr} (5s5p \; ^3P)$ asymptotes. The solid black lines denote $^{2}\Sigma^+$ states, the dashed red lines $^{2}\Pi$ states, the dotted blue line a $^{4}\Pi$ state and the dash-dotted green line a $^{4}\Sigma^+$ state.}
  \label{fgr:RbSr_pec_theory}
\end{figure}

The electronic states relevant to the present work are those dissociating into the two lowest atomic asymptotes $\mathrm{Rb} (5s\;^2S)+\mathrm{Sr} (5s^2\;^1S)$ and $\mathrm{Rb} (5p \; ^2P) + \mathrm{Sr} (5s^2 \; ^1S)$, see Fig.\,\ref{fgr:RbSr_pec_theory}. Our thermoluminescence spectra are dominated by transitions between levels belonging to the $\mathrm{X}(1)^2\Sigma^+$ and $\mathrm{B}(2)^2\Sigma^+$ states. Two-colour PA spectroscopy explores the $\mathrm{X}(1)^2\Sigma^+$ ground-state potential, by using intermediate molecular levels supported by potentials dissociating into the $\mathrm{Rb}(5s \; ^2S) + \mathrm{Sr} (5s5p \; ^3P)$ asymptote, see Fig.\,\ref{fgr:RbSr_pec_theory}. From our combined measurements we therefore derive quantitative information about the $\mathrm{X}(1)^2\Sigma^+$ and $\mathrm{B}(2)^2\Sigma^+$ states. Since for both states the projection $\Lambda$ of the electronic angular momentum on the internuclear axis is zero, spin-orbit coupling vanishes and Hund case (b) is the appropriate representation for the rotating molecule \cite{brown_carrington_2003}. The corresponding basis vectors are $|\Lambda,N,S,J\rangle$, where $N$ is the momentum given by the coupling between the corresponding angular momentum vector of $\Lambda$ and the nuclear orbital momentum, $S$ is the electron spin and $J$ is the total electronic angular momentum. Moreover, both the atomic and molecular levels are described by the total angular momentum of the Rb atom \cite{RbSrFFR}, labelled $F$ for the molecule and $f^{Rb}$ for the atom.

\section{PA spectroscopy of weakly-bound levels}
\label{sec:ColdExpPart}

In this section, we study the bound levels supported by the $\mathrm{X}(1)^2\Sigma^+$ RbSr ground-state potential close to the dissociation threshold using $\mu\mathrm{K}$ atomic clouds. We first describe the two-colour photoassociation spectroscopy we use to observe and characterize weakly-bound RbSr molecular levels. We then present the characteristics of the atomic mixture samples studied here. We give the spectroscopy results and discuss their uncertainties. We detail our data analysis and discuss the physical quantities that can readily be extracted from this type of data, such as the zero-energy semi-classical action and the van der Waals dispersion coefficients, which determine the spectrum of weakly-bound levels and the atomic scattering properties. We use our findings to confirm the identification and position of recently observed Fano-Feshbach resonances \cite{RbSrFFR}, and include these data into our analysis. Finally, we corroborate the overall analysis by comparing the \textit{s}-wave scattering lengths inferred by our model with the results of cross-thermalization measurements.

\subsection{Overview of two-colour photoassociation spectroscopy}
\begin{figure}[ht]
\centering
  \includegraphics[width=\columnwidth]{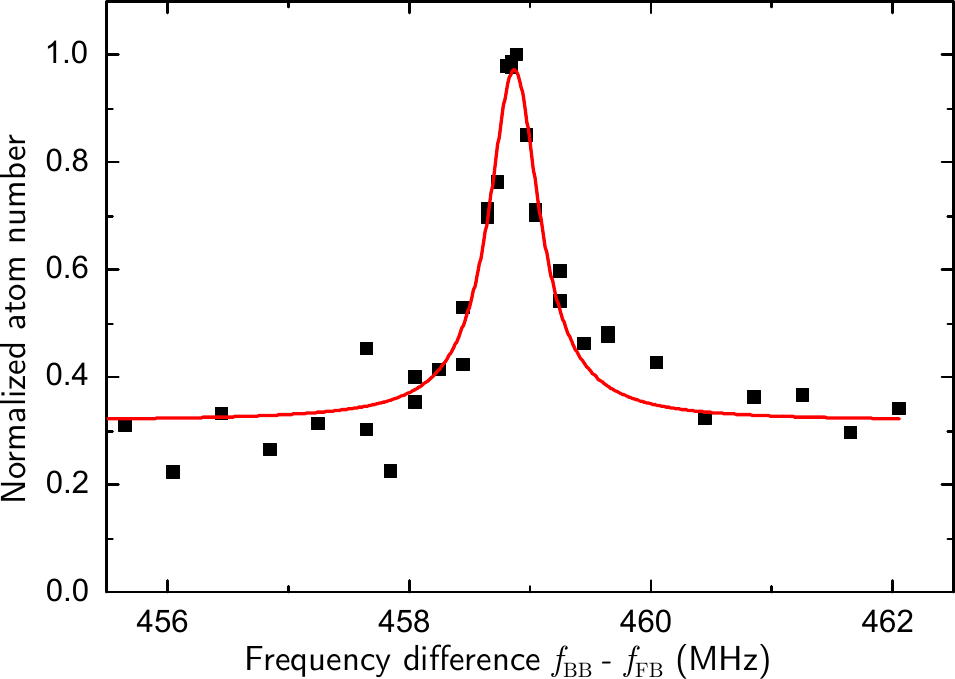}
  \caption{Typical two-colour photoassociation spectroscopy signal. The plot shows the number of Rb atoms in the $f^{\mathrm{Rb}}=1, \, m_f^{\mathrm{Rb}}=1$ level normalized to the atom number in the other two $f^{\mathrm{Rb}}=1, \, m_f^{\mathrm{Rb}}$ levels, as function of the frequency difference between $L_{\mathrm{BB}}$ and $L_{\mathrm{FB}}$, recorded during a scan of the $L_{\mathrm{BB}}$ frequency, while $L_{\mathrm{FB}}$ is on resonance with an $a$ to $e$ transition. This signal corresponds to the $\{ \nu=66,\, N=0,\, F=1 \}$ level of the $^{87}\mathrm{Rb}$-$^{88}\mathrm{Sr}$ ground-state potential, see Table\,\ref{tbl:WeaklyBoundSpectrumLJ}.}
  \label{fgr:2colorSignal}
\end{figure}

We carry out two-colour PA spectroscopy to observe weakly-bound $\mathrm{X}(1)^2\Sigma^+$ levels and measure their energies referenced to the energy of the atomic scattering state Rb$(^2S_{1/2},f^{\mathrm{Rb}}=1)$+Sr$(^1S_0)$. Two-colour PA spectroscopy exploits the presence of an optically-excited molecular level $e$, which is coupled to an atom-pair state $a$ by the free-bound laser $L_{\mathrm{FB}}$ with frequency $f_{\mathrm{FB}}$. When this laser is resonant with the $a$ to $e$ transition, pairs of colliding atoms $a$ are transferred to $e$, from where they spontaneously decay to low-lying molecular levels, resulting in atom loss \footnote[1]{We confirm that such loss originates from the formation of RbSr molecules and not $\mathrm{Rb}_2$ or $\mathrm{Sr}_2$ molecules, by verifying that the loss only occurs if both elements are present.}. If an additional bound-bound laser $L_{\mathrm{BB}}$ with frequency $f_{\mathrm{BB}}$ is tuned on resonance with a molecular transition between $e$ and a weakly-bound molecular level $m$ of the $\mathrm{X}(1)^2\Sigma^+$ ground state, a significant light shift pushes $e$ out of resonance with $L_{\mathrm{FB}}$. The loss induced by $L_{\mathrm{FB}}$ is then suppressed, resulting in an atom number peak when varying the frequency of $L_{\mathrm{BB}}$, see the example in Fig.\,\ref{fgr:2colorSignal}. The energy $E$ of the molecular level $m$ referenced to the energy of the atom pair $a$ is directly given at this peak by $E = h \times (f_{\mathrm{BB}} - f_{\mathrm{FB}})$, where $h$ is the Planck constant. In the limit of low temperature and small external fields, the molecular binding energy $E_b$ is equal to $E$ for levels with $F = 1$, and $E_b = E + E_{\mathrm{hf}}$ for levels with $F = 2$, where $E_{\mathrm{hf}}$ is the Rb hyperfine splitting.

In order to detect weakly-bound levels up to the least bound ones, we exploit similarly weakly-bound levels supported by the electronically excited potentials correlating to the $\mathrm{Rb}(5s \; ^2S_{1/2}) + \mathrm{Sr} (5s5p \; ^3P_1)$ asymptote \cite{CiameiThesis, PiotrOlivierRbSr}, see Fig.\,\ref{fgr:RbSr_pec_theory}. These levels provide sufficient Franck-Condon factors between $e$ and $m$, while the narrow linewidth of the nearby Sr intercombination transition results only in small losses and heating by off-resonant scattering of photons on Sr atoms.

\subsection{Sample conditions and spectroscopy setup}
We prepare the desired ultracold mixtures of $\mathrm{Rb}\operatorname{-}\mathrm{Sr}$ isotopes as in our previous works \cite{RbSrSample, BayerleThesis, RbSrFFR}. We keep the mixture in a crossed-beam dipole trap formed by one 1064-nm horizontal elliptical beam with a vertical waist of $19(1)\,\mu\mathrm{m}$ and a horizontal waist of $313(16)\,\mu\mathrm{m}$, and one 1064-nm (or 1070-nm) vertical beam with a waist of $78(2)\,\mu\mathrm{m}$ (or $90(5)\,\mu\mathrm{m}$). When studying $^{87}$Sr, we add a 532-nm horizontal beam with vertical (horizontal) waist of $19(1)\,\mu\mathrm{m}$ ($219(4)\,\mu\mathrm{m}$) to increase the trap depth, in order to capture more Sr atoms. We prepare $^{87}$Rb in its hyperfine ground level $(^2S_{1/2}, f^{\mathrm{Rb}}=1)$ with almost equal population of the Zeeman sub-levels $m^{\mathrm{Rb}}_f=0,\pm 1$. During PA spectroscopy, we measure each $m^{\mathrm{Rb}}_f$ population separately via time-of-flight expansion in a Stern-Gerlach magnetic field gradient. Bosonic Sr isotopes $^{84}$Sr and $^{88}$Sr have zero nuclear magnetic moment leading to a structureless $^1S_0$ ground state. Fermionic $^{87}$Sr has a non-zero nuclear magnetic moment of $i^{\mathrm{Sr}}=9/2$ and is prepared in the stretched level $m^{\mathrm{Sr}}_i=9/2$ or $m^{\mathrm{Sr}}_i=-9/2$ via optical pumping.

The atomic samples used to study $^{87}\mathrm{Rb}^{84}\mathrm{Sr}$ and $^{87}\mathrm{Rb}^{88}\mathrm{Sr}$ molecules have a temperature of $1.0(1)\,\mu\mathrm{K}$, average densities of $0.3-4 \times 10^{12}\,\mathrm{cm}^{-3}$ for Sr and $2-7 \times 10^{12}\,\mathrm{cm}^{-3}$ for Rb (summing over all $m^{\mathrm{Rb}}_f$ levels), and the trap frequencies are $\left\lbrace \omega_x,\, \omega_y,\,\omega_z \right\rbrace=2\pi \times \left\lbrace 66(6),\, 57(6),\, 560(50) \right\rbrace\,\mathrm{Hz}$ for Sr and $2\pi \times \left\lbrace 110(10),\, 95(9),\, 950(80) \right\rbrace\,\mathrm{Hz}$ for Rb \footnote[1]{The error bars on the trap frequencies are dominated by the uncertainty on the waists previously stated.}, where the $z$-axis is vertical \footnote[2]{The sample used to detect the $^{87}\mathrm{Rb}^{88}\mathrm{Sr}$ level at a binding energy of $h \times 459\,\mathrm{MHz}$ has a temperature of $1.5(1)\,\mu\mathrm{K}$, average densities $6.4(2.2) \times 10^{12}\,\mathrm{cm}^{-3}$ for Sr and $8.4(3.0) \times 10^{11}\,\mathrm{cm}^{-3}$ for Rb, and trapping frequencies $\left\lbrace \omega_x,\, \omega_y,\, \omega_z \right\rbrace=2\pi \times \left\lbrace 57(7),\, 8(1),\, 930(100) \right\rbrace\,\mathrm{Hz}$ for Sr and $2\pi \times \left\lbrace 97(10),\, 14(2),\, 1600(200) \right\rbrace\,\mathrm{Hz}$ for Rb.}. The samples used to study the fermionic $^{87}$Rb$^{87}$Sr molecules have a temperature of $1.5(1)\,\mu\mathrm{K}$, average densities in the range $2-6 \times 10^{11}\,\mathrm{cm}^{-3}$ for Sr and $0.8-3 \times 10^{12}\,\mathrm{cm}^{-3}$ for Rb, and trapping frequencies in the range $\left\lbrace \omega_x,\, \omega_y,\, \omega_z \right\rbrace = 2\pi \times \left\lbrace 70-80,\, 55-70,\, 590-640 \right\rbrace\,\mathrm{Hz}$ for Sr and $2\pi \times \left\lbrace 110-130,\, 95-125,\, 820-930 \right\rbrace\,\mathrm{Hz}$ for Rb.

The PA beam, containing both $L_{\mathrm{FB}}$ and $L_{\mathrm{BB}}$, propagates horizontally at a $\sim 30^{\circ}$ angle from the axis of the horizontal dipole trap and has a waist of either $60(1)\,\mu\mathrm{m}$ or $110(10)\,\mu\mathrm{m}$, depending on the transition strength and the available laser power. $L_{\mathrm{FB}}$ and $L_{\mathrm{BB}}$ are derived from the same master oscillator, either via injection-lock or beat-lock, which ensures good coherence between them (typically below $100\,\mathrm{Hz}$ for injection-locked and $\sim 30\,\mathrm{kHz}$ for beat-locked lasers). We apply a homogeneous magnetic field in the range of 0 to $20\,\mathrm{G}$. We vary the polarization and frequency of $L_{\mathrm{FB,BB}}$ as required to optimally detect a specific molecular level. We adjust the pulse time and power of $L_{\mathrm{FB}}$ in order to induce 70 to $90\,\%$ loss of Sr atoms. The $L_{\mathrm{BB}}$ intensity is chosen such as to obtain a good signal-to-noise ratio for the two-colour PA signal, while not being limited by off-resonant scattering of photons on Sr atoms.

\subsection{Experimental results}
We have observed a total of 10 molecular levels via two-colour PA spectroscopy, of which we report the energies $E$ in Table\,\ref{tbl:PA results}. We also report the one-colour PA lines used, the type of transition induced, the angular momentum projections involved and the bound-bound Rabi frequency, if characterized. Levels with negative values of $E$ are necessarily levels with $F=2$. The typical error of $E_b$ is significantly larger than the uncertainty of $h \times (f_{\mathrm{BB}} - f_{\mathrm{FB}})$, and is the result of several sources of uncertainty.

\begin{table*}[ht]
	\small
  	\caption{Results of two-colour PA spectroscopy. The energy $E$ is given by the two-colour frequency detuning $E /h=\nu_{L_{\mathrm{BB}}}-\nu_{L_{\mathrm{FB}}}$ at $B=0\,\mathrm{G}$. $\Delta$ is the detuning of $L_{\mathrm{FB}}$ from the $\mathrm{Sr}$ $^1\mathrm{S}_0 - ^3\mathrm{P}_1$ transition for bosonic molecules and from the $\mathrm{Sr}$ $(^1\mathrm{S}_0, f^{\mathrm{Sr}}=9/2) - (^3\mathrm{P}_1, f^{\mathrm{Sr}}=11/2)$ transition for $^{87}\mathrm{Rb}^{87}\mathrm{Sr}$ molecules. The column labelled ``Transition'' shows the type of transition addressed by both $L_{\mathrm{FB}}$ and $L_{\mathrm{BB}}$. The labels $m_f^{at}$ and $m_F^{mol}$ are the projections of the total angular momentum on the quantization axis, neglecting nuclear rotation and the nuclear spin of $^{87}\mathrm{Sr}$, for the atomic and molecular levels, respectively. For some excited levels $m_F^{mol,e}$ is not known. Such levels are all high-field seeking. The last column gives the bound-bound Rabi frequencies, if measured, between molecular levels $m$ and $e$ in the electronic ground and excited states}
\begin{tabular*}{\textwidth}{@{\extracolsep{\fill}}lllccl}
    \hline
Isotopologue & $E/h$ (MHz) & $\Delta$ (MHz) & Transition & $m_f^{at}$, $m_F^{mol,e}$, $m_F^{mol,m}$ & $\Omega/2\pi\,(\mathrm{kHz}/\sqrt{\mathrm{mW/cm^2}})$\\
\hline
\noalign{\smallskip}
$^{87}\mathrm{Rb}^{84}\mathrm{Sr}$ & $29.01(3)$ & $173.5(2)$, $427.8(2)$ & $\pi$ & $0$, $0$, $0$ & -, $16(1)$\\
 & $744.53(3)$ & $173.5(2)$, $427.8(2)$ & $\pi$ & $0$, $0$, $0$ & $6.7(3)$, $285(10)$\\
\hline
\noalign{\smallskip}
$^{87}\mathrm{Rb}^{87}\mathrm{Sr}$ & $199.97(17)$ & $686.79(23)$ & $\sigma^-$ & $-1$,-,$-1$ & $6.0(5)$\\
 & $287.27(18)$ & $686.79(23)$ & $\sigma^-$ & $-1$, -, $-1$ & $22.2(2.0)$\\
 & $1950.24(11)$ & $686.79(23)$ & $\sigma^-$ & $-1$, -, $-1$ & -\\
\hline
\noalign{\smallskip}
$^{87}\mathrm{Rb}^{88}\mathrm{Sr}$ & $-6476.80(4)$ & $41.39(60)$ & $\sigma^{\pm}$ & $0$, -, $0$ & $0.31(6)$\\
 & $-4677.78(15)$ & $260.54(5)$ & $\sigma^{\pm}$ & $0$, $0$, $0$ & -\\
 & $356.99(3)$ & $41.39(60)$ & $\pi$ & $0$, -, $0$ & $3.39(25)$\\
 & $458.90(22)$ & $53.5(4)$ & $\pi$ & $1$, -, $1$ & $3.1(1.3)$\\
 & $2153.83(15)$ & $260.54(5)$ & $\pi$ & $0$, $0$, $0$ & $0.59(13)$\\
\hline
\label{tbl:PA results}
\end{tabular*}
\end{table*}

\begin{figure}[ht]
\centering
  \includegraphics[width=\columnwidth]{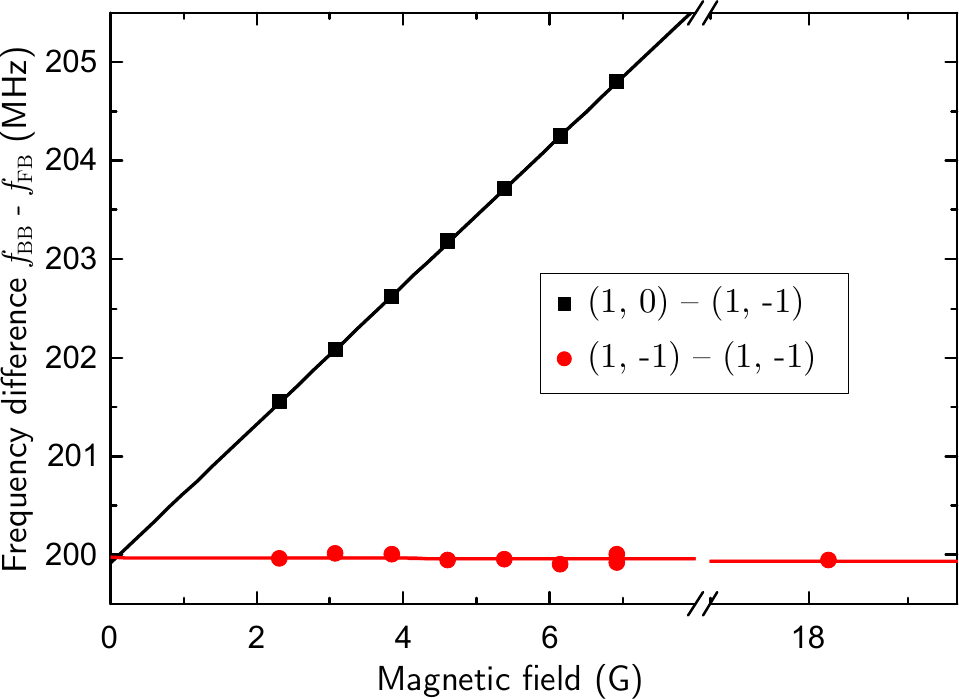}
  \caption{Molecular level energy $E = h \times (f_{\mathrm{BB}}-f_{\mathrm{FB}})$ as function of the magnetic field. The $({F, \, m_F}) - (f^{\mathrm{Rb}}, \, m_f^{\mathrm{Rb}} )$ transitions $(1,0) - (1,-1)$ (black squares) and $(1,-1) - (1,-1)$ (red disks) are shown together with linear fits. The error bars of the measurements are smaller than the symbol sizes.}
  \label{fgr:GroundStateMagneticMoment}
\end{figure}

The first significant error contribution comes from the differential Zeeman shift between the atom-pair level $a$ and the molecular level $m$. In order to minimize this contribution, we exploit the fact that, in the case of equal spin quantum numbers $F=f^{\mathrm{Rb}}$ and $m_F=m_f^{\mathrm{Rb}}$, this shift is vanishingly small for weak binding energy of the molecular level $m$, see the example of Fig.\,\ref{fgr:GroundStateMagneticMoment}. We thus drive two-colour transitions between the atom pair in $f^{\mathrm{Rb}}=1$ and molecular levels with $F=1$ and $m_F=m_f^{\mathrm{Rb}}$. For the example of Fig.\,\ref{fgr:GroundStateMagneticMoment} with $m_F=m_f^{\mathrm{Rb}}$, we derive a small differential magnetic moment of $-2.0(2.0)\,\mathrm{kHz/G}$. Such shift extrapolated to all measured points results in a maximum systematic shift in the range $0.2-20.0\,\mathrm{kHz}$. For molecular levels with $F=2$, we drive magnetically insensitive two-colour transitions with $m_F = m_f^{\mathrm{Rb}}= 0$. This results in a systematic shift of at the most $10\,\mathrm{Hz}$ for the measured points.

The second error contribution is the light shift on the two-colour transition arising from the spectroscopy lasers themselves. We have characterized it for some of the points and estimated it for the others as explained in our previous work \cite{EfficientSr2}. This amounts to shifts up to $100\,\mathrm{kHz}$ for the typical laser powers used here.

The third error contribution is the light shift on the two-colour transition arising from the dipole trap. Similarly to the previous error contribution, it is characterized for some of the points, and estimated for the others. Given the trap used here, only the differential polarizability between atoms and molecules affecting the centre-of-mass Hamiltonian is relevant. For $\mathrm{Sr}_2$ we have shown that the relative variation of the polarizability is smaller than $1\,\%$ \cite{EfficientSr2}. Here we use the conservative value of $5\,\%$ to estimate the errors if not characterized.

The last error contribution comes from thermal shifts and is typically negligible, as a temperature around $1\,\mu\mathrm{K}$ corresponds to an energy of about $h \times 20\,\mathrm{kHz}$.

\subsection{Data analysis}
\label{subsec:ColdExpPart.DataAnalysis}

\subsubsection{Line attribution and estimation of physical quantities.}
\label{Line attribution and estimation of physical quantities}
The first step in a quantitative analysis of the weakly-bound spectrum probed by two-colour photoassociation is a line attribution, by which we mean the assignment of quantum numbers to both atomic and molecular levels used in the measurements of $E$. In particular, the angular momenta $F$ and $N$ of molecular levels are not known, and their proper assignment is crucial to the success of any model-fitting attempt. The angular momenta of atomic levels are partially known: we measure the Rb atomic spin angular momentum $f^{\mathrm{Rb}}=1$ and its projection $m_f$, and for $^{87}\mathrm{Sr}$ we measure the nuclear momentum projection $m^{\mathrm{Sr}}_i$. However, despite the low temperature of the sample, the atom-pair orbital momentum is not known, but will be zero for the majority of cases. The possible molecular angular momenta can be restricted by considering the atomic angular momenta and the changes of angular momentum allowed by two-colour PA. All possible assignments of quantum numbers must then be tested by the fit in order to find the best one.

Attributing quantum numbers to molecular levels is not trivial, especially in cases of sparse spectral data like the present one. In the spectral region considered here, i.e. for $E_b<<D_e$ with $D_e$ the molecular potential depth, only universal, model-independent properties are invoked in order to attribute quantum numbers. These properties are:
\begin{enumerate}
\item the asymptotic behaviour of the interaction potential $V_g$ as $V_g(r)\rightarrow -C_6/r^6$ at large internuclear distance $r\gg R_L$, where $R_L$ is the LeRoy radius;
\item the presence of a strong repulsive wall at the inner classical turning point, i.e. $|\dfrac{dV_g}{dr}(R_1)|\gg |\dfrac{dV_g}{dr}(R_2)|$, where $R_1$ and $R_2$ are the inner and outer classical turning points for vibrational motion;
\item mass-scalability under the Born-Oppenheimer approximation, due to the presence of a single electronic state. 
\end{enumerate}
Property 1 implies that the semi-classical phase accumulation of the zero-energy scattering wavefunction $\Phi^0_{WKB}=\Phi_{WKB}(E_b=0)$ is well defined. Moreover, because of both property 2 and the condition $E_b<<D_e$, $\Phi_{WKB}$ is to a large extent model-independent \cite{AnalyticsFlambaum2}. Property 3 implies that a single-channel Hamiltonian, containing $V_g(r)$, explains all spectral data for different isotopologues via simple mass-scaling. Based on these considerations, a single well-defined value $\Phi^0_{WKB}$ referenced to an isotopologue of choice is sufficient to describe our system. In this work we choose the isotopologue with highest abundance $^{85}\mathrm{Rb}^{88}\mathrm{Sr}$ as our reference. As a consequence of these properties, we expect that two physical quantities $\Phi^0_{WKB}$ and $C_6$ can be extracted from our data by fitting our complete dataset and assigning the quantum numbers consistently. We employ a simple semi-classical approach to find the correct attribution of quantum numbers $F$ and $N$ \cite{AKAKE-I}.

We use this fitting strategy on the two-colour photoassociation spectroscopy data presented in Table\,\ref{tbl:WeaklyBoundSpectrumLJ} (labelled as ``PA'' in the ``Method'' column). Only a single attribution of quantum numbers delivers a satisfactory fit, which we report in Table\,\ref{tbl:WeaklyBoundSpectrumLJ} in columns $F$ and $N$. This simple fitting strategy is thus sufficient to provide an unambiguous attribution for these two quantum numbers, while the attribution of $\nu$ still presents some uncertainty. As expected, the vast majority of observed molecular levels are either $N=0$ or $N=2$. Even so, let us note the presence of two $N=3$ levels for $^{87}\mathrm{Rb}^{88}\mathrm{Sr}$, which might seem inconsistent with our ultracold sample temperature. However, due to the presence of a virtual near-threshold level in this isotopologue, the amplitude of the \textit{p}-wave scattering wavefunction at the one-colour PA Condon point is less than a factor of 3 smaller than that of typical \textit{s}-wave scattering states. The fit also provides a first estimation for $\Phi^0_{WKB}$ and $C_6$. We extract the zero-energy semi-classical action $\Phi^0_{WKB}=67.42(1)$, which gives 67 bound levels for $^{87}\mathrm{Rb}^{84}\mathrm{Sr}$ and $^{87}\mathrm{Rb}^{87}\mathrm{Sr}$, and 68 bound levels for $^{87}\mathrm{Rb}^{88}\mathrm{Sr}$, and we extract the dispersion coefficient $C_6=1.78(2)\times 10^7\,$\r{A}$^6\,\mathrm{cm}^{-1}$. Both quantities are determined with better accuracy and precision in the following sections.

\begin{table*}[ht]
\small
\caption{List of observed molecular levels, with the experimentally determined binding energies $E_{b}$ and corresponding errors approximated by the theoretically estimated shift from the variation of the hyperfine splitting $\delta E_{\mathrm{hf}}$. The values $E_{b}^{th}$ represent the binding energies given by the best fit in Sec.\,\ref{sec:inclusionFFR}. The quantum numbers $\{\nu,\,N,\,F\}$ identifying each level are also shown. The vibrational quantum number $\nu$ is counted starting from the lowest level, which has $\nu=0$. The ``Method'' column specifies whether the levels are determined by two-colour photoassociation (PA) or Fano-Feshbach (FFR) spectroscopy, or both}
\begin{tabular*}{\textwidth}{@{\extracolsep{\fill}}lllllllll}
\hline
Isotopologue & $E_{b}/h \, \mathrm{(MHz)}$ & $\delta E_{\mathrm{hf}} /h \, \mathrm{(MHz)}$ & $E_{b}^{th} /h \, \mathrm{(MHz)}$ & $\nu$ & $N$ & $F$ & Method\\
\hline
\noalign{\smallskip}
$^{87}\mathrm{Rb}^{84}\mathrm{Sr}$ & $29.01(3)$ & $0.094$ & $28.93$ & $66$ & $0$ & $1$ & PA\\
 & $744.53(3)$ & $0.82$ & $745.27$ & $65$ & $0$ & $1$ & PA\\
\hline
\noalign{\smallskip}
$^{87}\mathrm{Rb}^{87}\mathrm{Sr}$ & $199.97(17)$  & $0.34$ & $199.90$ & $66$ & $2$ & $1$ & PA,FFR\\
 & $287.27(18)^{a}$ & $0.44$ & $287.30$ & $66$ & $0$ & $1$ & PA,FFR\\
 & $288.2(4)^{a}$ & $0.44$ & $287.30$ & $66$ & $0$ & $2$ & FFR\\
 & $1950.24(11)$  & $1.56$ & $1953.00$ & $65$ & $0$ & $1$ & PA\\
 & $5991.8(1.4)$  & $3.30$ & $5991.64$ & $64$ & $2$ & $2$ & FFR\\
 & $6233.8(1.0)$ & $3.39$ & $6232.14$ & $64$ & $0$ & $2$ & FFR\\
\hline
\noalign{\smallskip}
$^{87}\mathrm{Rb}^{88}\mathrm{Sr}$ & $356.99(3)^{a,b}$ & $0.50$ & $357.21$ & $66$ & $2$ & $1$ & PA\\
 & $357.87(4)^{a,b}$ & $0.50$ & $357.21$ & $66$ & $2$ & $2$ & PA\\
 & $458.90(22)$ & $0.59$ & $459.12$ & $66$ & $0$ & $1$ & PA\\
 & $2153.83(15)^{a,b}$ & $1.67$ & $2158.40$ & $65$ & $3$ & $1$ & PA\\
 & $2156.91(15)^{a,b}$ & $1.67$ & $2158.40$ & $65$ & $3$ & $2$ & PA\\
 & $7401.01(66)$ & $3.80$ & $7397.47$ & $64$ & $0$ & $2$ & FFR\\
\hline
\label{tbl:WeaklyBoundSpectrumLJ}  
\end{tabular*}
$^{a}$ For the fit, we use the mean value of each pair.\\
$^{b}$ We use the measured energies of these pairs to estimate $\delta E_{\mathrm{hf}}$.\\
\end{table*}

\subsubsection{Extraction of physical quantities.}
\label{sec:PrecisionFitPA}
Based on the unambiguous quantum number attribution of $F$ and $N$ explained in the previous section, we check the consistency of our data with the universal long-range dispersion and extract the relevant physical quantities from a fit of a second model. We assess consistency with our data using the reduced chi-square as figure of merit:
\begin{equation}
\label{Chi-square}
 \widetilde{\chi}^2 =\frac{1}{\mathrm{DOF}}\times \sum_{i}\left( {\frac{E_{b,i}^{exp}-E_{b,i}^{th}}{\Delta E_{b,i}^{exp}}} \right)^2,
\end{equation}
where $\mathrm{DOF}$ is the number of degrees of freedom in the fit \footnote[1]{The $\mathrm{DOF}$ are defined as the number of experimental data points minus the number of fit parameters.}, $i$ runs over the experimentally observed levels, $E_{b,i}^{exp}$ is the $i^{\mathrm{th}}$ measured binding energy (BE), $E_{b,i}^{th}$ is the $i^{\mathrm{th}}$ predicted BE and $\Delta E_{b,i}^{exp}$ is the experimental error of $E_{b,i}^{exp}$. We consider $\widetilde{\chi}^2$ to be good if close to unity, i.e. $\widetilde{\chi}^2 \simeq 1$. Since probability levels can only be associated to confidence intervals if the experimental error distribution is known, we only state confidence intervals based on a given absolute variation of $\widetilde{\chi}^2$, without quantitative knowledge of the associated probability level.

We fit a single-channel Hamiltonian model to our experimental data \cite{AKAKE-I} in order to retrieve the relevant physical information, i.e. the zero-energy semi-classical action and dispersion coefficients. This means we require consistency between BEs measured via two-colour PA spectroscopy and the bound spectrum supported by the Hamiltonian
\begin{equation}
\label{HamiltonianModel}
 \widetilde{H}=\widetilde{T}+\widetilde{V}_{int}+\widetilde{V}_{rot}=-\frac{\hbar^2}{2\mu}\frac{d^2}{dr^2}+V_{g}(r)+\frac{\hbar^2}{2\mu}\frac{N(N+1)}{r^2},
\end{equation}
where $\widetilde{T}$ is the kinetic energy operator, $\mu$ is the reduced mass, $\widetilde{V}_{rot}$ is the rotational energy operator and $\widetilde{V}_{int}$ is the interaction operator corresponding for the ground state to $V_{g}$, which obeys the properties enumerated in Sec.\,\ref{Line attribution and estimation of physical quantities}. For simplicity, we here use the generalized Lennard-Jones model for $V_{g}$:
\begin{equation}
\label{Lennard_Jones}
 V_{LJ}(r)=\frac{C_6}{r^6}\times \left( \left( \frac{\sigma}{r} \right)^6 -1 \right)-\sum\limits_{n\geq 2}^{N_{vdW}} \frac{C_{2(2+n)}}{r^{2(2+n)}},
\end{equation}
which contains the leading order dispersion coefficients $C_6, C_8,...,C_{2(2+N_{vdW})}$. The maximum order $N_{vdW}$ used in the long-range asymptotic expansion is chosen as the lowest number that is able to provide a good fit of our data by the weakly-bound spectrum supported by $V_{LJ}$. The parameter $\sigma$ is used to tune the short-range phase accumulation.

Due to the presence of a single electronic ground state in RbSr, the simple single-channel model (\ref{HamiltonianModel}) with the potential of eqn\,(\ref{Lennard_Jones}) is sufficient to provide a unique attribution of the quantum numbers $F$, $N$ and $\nu$ for our experimental data. However, it is in general not sufficient to fit high-resolution spectra to experimental accuracy. This is mostly due to the fact that the two $^2\Sigma^+$ PECs of $F=1$ and $F=2$ character are not exactly parallel \cite{PiotrRbSrMagnetoassociation}. At large internuclear separation $r$ the splitting $E_{\mathrm{hf}}(r)$ between these PECs is the Rb atom hyperfine splitting, whereas it is reduced by about $10\,\%$ at the bottom of the PECs. This effect is due to the reduction of the electronic density at the Rb nucleus because of the bonding with Sr. Although extremely small, it is responsible for the strongest Fano-Feshbach resonances recently observed in RbSr \cite{RbSrFFR}. In the present work, our precision and accuracy are enough to reveal hints for this effect, appearing as significant differences in the BEs of levels with the same $\nu$ and $N$ quantum numbers but different $F$, see the pairs of BEs of $^{87}$Rb$^{88}$Sr $\{ \nu=66, \, N=2 \}$ and $\{ \nu=65, \, N=3\}$ in Table\,\ref{tbl:WeaklyBoundSpectrumLJ}. However, our data are not sufficient to extract this shift reliably and include it in our model \footnote[1]{The two observed shifts mentioned involve rotationally excited molecular levels with unknown spin-rotation coupling, hence they do not directly yield the shift under discussion. The simplest experiment able to characterize this shift requires the measurement of pairs of rotationless levels at different BEs.}. We therefore keep a single-channel model and take this effect into account as a systematic error contribution to $\Delta E_{b,i}$. This contribution is estimated using the aforementioned differences in BEs and knowing that the change in hyperfine splitting scales as $\delta E_{\mathrm{hf}} \propto E_{b}^{2/3}$ close to the dissociation threshold \cite{Brue2013PRAFFR}. These estimated shifts $\delta E_{\mathrm{hf},i}$, which dominate the errors $\Delta E_{b,i}$, are of the same order of magnitude as the shift predictions from \textit{ab-initio} results \cite{PiotrRbSrMagnetoassociation}, and are listed in Table\,\ref{tbl:WeaklyBoundSpectrumLJ}. When BEs of both hyperfine states are measured, the mean binding energy is used in the fit, see Table\,\ref{tbl:WeaklyBoundSpectrumLJ}. The quality of this estimation is assessed a posteriori via the $\widetilde{\chi}^2$ of the best fit, labelled $\widetilde{\chi}_{\mathrm{min}}^2$.

\begin{figure}[ht]
\centering
  \includegraphics[width=\columnwidth]{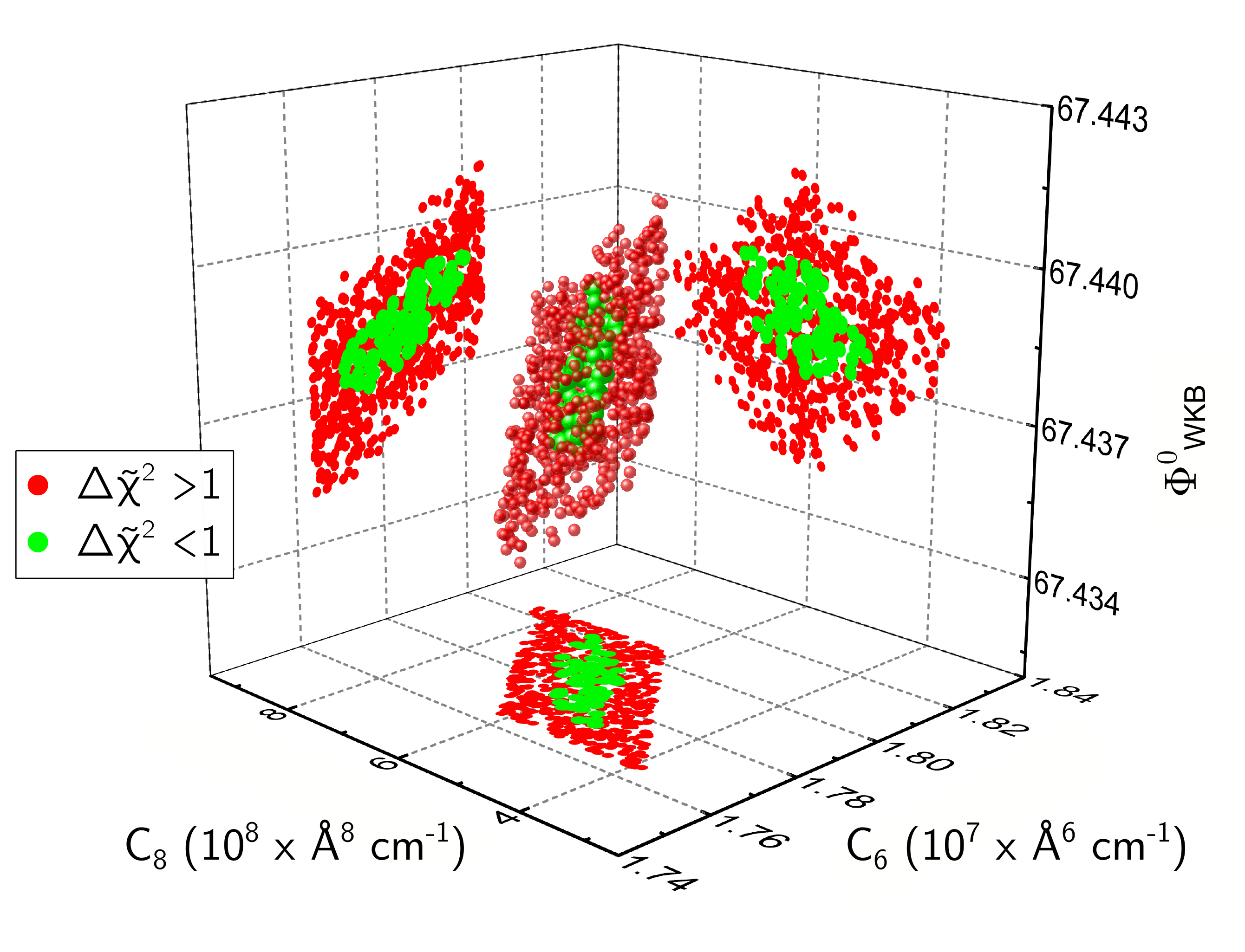}
  \caption{Confidence intervals. The plot shows $\widetilde{\chi}^2$ for configurations sampled close to the best fit configuration in the fit-parameter space, with its projections on the 2D coordinate planes. The total confidence regions corresponding to $\Delta\widetilde{\chi}^2 <1$, used for error estimates, are indicated as green balls in the centre.}
  \label{fgr:AnalysisOnFITResults}
\end{figure}

We fit the model Hamiltonian (\ref{HamiltonianModel}) to our PA spectroscopy data, using $\{ \sigma,C_{6},...,C_{2(2+N_{vdW})}\}$ as independent fit parameters and we retrieve the zero-energy semi-classical action $\Phi^0_{WKB}=\Phi^0_{WKB}(\sigma,C_{6},...,,C_{2(2+N_{vdW})})$. For $N_{vdW}=1,\,2,\,3$ we obtain for the best fits $\widetilde{\chi}_{\mathrm{min}}^2=41,\,0.24,\,0.32$, respectively. This shows that the inclusion of $C_6$ and $C_8$ terms is \textit{necessary} and \textit{sufficient} to model our data. We obtain the best fit parameters $\sigma=5.012941656601387\,$\r{A}, $C_6 = 1.784438900566861\times 10^7\,$\r{A}$^6\,\mathrm{cm}^{-1}$, $C_8=6.18126306008073 \times 10^8\,$\r{A}$^8\,\mathrm{cm}^{-1}$ with $\mathrm{DOF} = 5$. The fit returns the physical quantities $C_6 = 1.784(15)\times 10^7\,$\r{A}$^6\,\mathrm{cm}^{-1}$, $C_8= 6.2(1.1) \times 10^8\,$\r{A}$^8\,\mathrm{cm}^{-1}$, and a corresponding $\Phi^0_{WKB}=67.4379(12)$ \footnote[1]{The number of vibrational levels is determined without uncertainty. All isotopologues have 67 vibrational levels, except for the two with the highest mass $^{87}\mathrm{Rb}^{88}\mathrm{Sr}$ and $^{87}\mathrm{Rb}^{87}\mathrm{Sr}$, which have 68.}. The errors stated in brackets correspond to, somewhat arbitrarily, the joint confidence region with $\Delta \widetilde{\chi}^2=\widetilde{\chi}^2-\widetilde{\chi}_{\mathrm{min}}^2=1$. In Fig.\,\ref{fgr:AnalysisOnFITResults} we show the configurations sampled by the fitting procedure that provide the evaluation of the confidence regions. The dispersion coefficients are consistent with theoretical predictions \cite{CICPDispersionCoeffMitroy, MBPTDispersionCoeffDEREVIANKO, DispersionCoeffStandard, PiotrRbSrMagnetoassociation}.

\begin{figure}[ht]
\centering
  \includegraphics[width=\columnwidth]{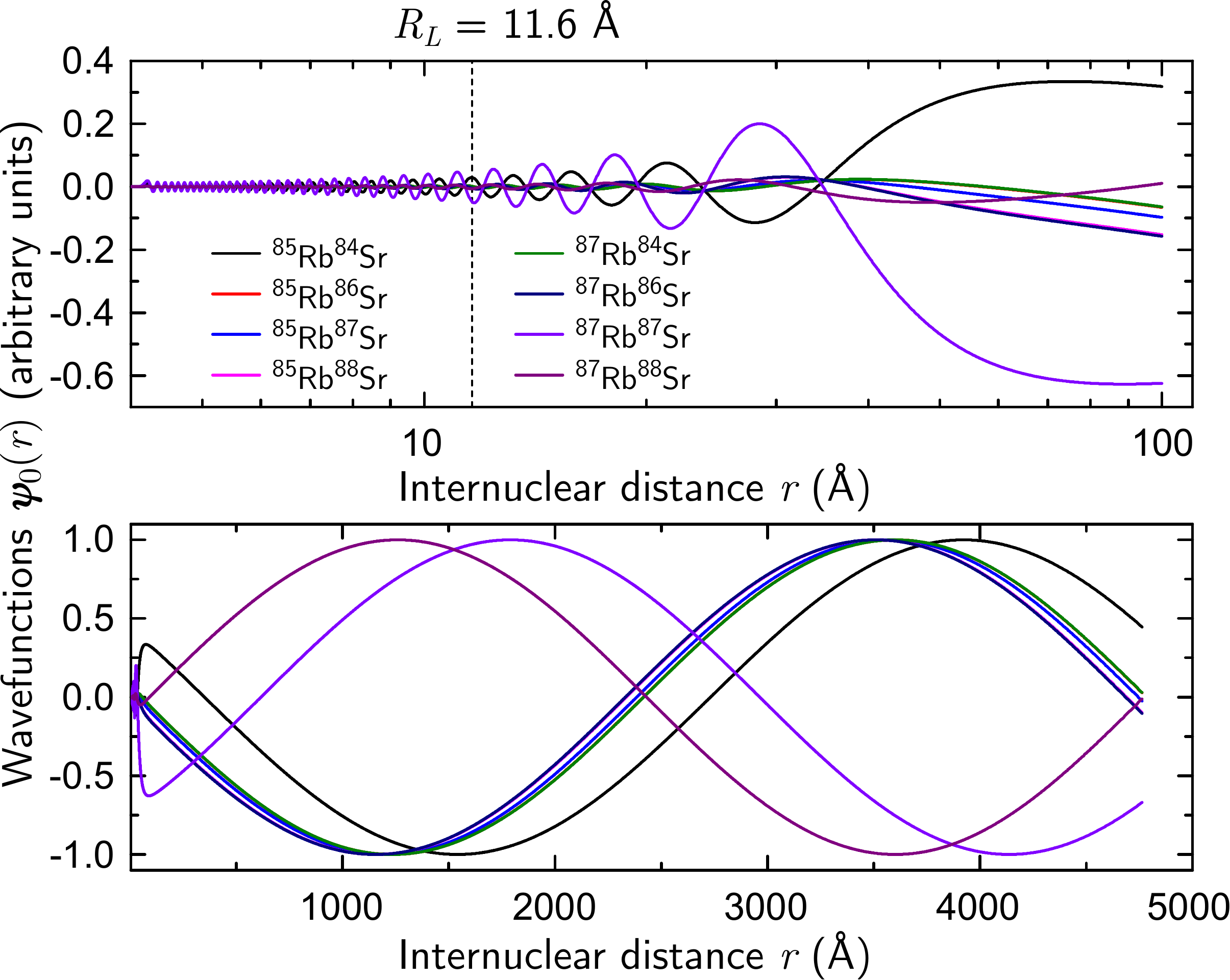}
  \caption{Atom-pair scattering wavefunctions. The top panel shows the scattering wavefunctions as function of the internuclear distance for all $\mathrm{Rb}\operatorname{-}\mathrm{Sr}$ isotopes in the range of typical PA Condon points, which are only meaningful for $r>R_L$. The bottom panel shows the scattering wavefunctions at large distance where the phase shift encodes the short-range physics. We note that two pairs of isotopes have incidentally almost the same reduced mass, hence almost overlapping wavefunctions.}
  \label{fgr:ScatteringWavefunctions}
\end{figure}

Atomic scattering properties at a given collisional energy and in the absence of external magnetic fields can be directly derived from the fitted PEC \cite{AKAKE-I}. Scattering wavefunctions are obtained by integration of the nuclear Schr\"{o}dinger equation for the appropriate angular momentum $N$, i.e. $\widetilde{H}(N) \psi_N(r) = E_{coll} \psi_N(r)$, where $\widetilde{H}(N)$ is the fitted Hamiltonian (\ref{HamiltonianModel}) with explicit $N$-dependence and $E_{coll}=\hbar^2 k^2/2 \mu$ is the collisional energy with wavevector $k$. Of particular interest for the cold atoms community are the scattering properties in the limit $E_{coll}\rightarrow 0$, which are dominated by \textit{s}-wave scattering, i.e. $N=0$. In this limit the scattering phase shift $\delta\phi \rightarrow -k a_s$  and the cross-section $\sigma_{s}\rightarrow 4\pi a_s^2$ are determined by a single parameter, the \textit{s}-wave scattering length $a_s$, which we report in Table\,\ref{tbl:SwaveScatteringLengths} for all stable isotopologues of RbSr \footnote[2]{The \textit{s}-wave scattering length $a_s$ is calculated by fitting $\psi_0$ at large $r$ with the known asymptotic behaviour $\psi_0(r) \rightarrow \alpha r+\beta$, where $\alpha$ and $\beta$ are fit parameters. $a_s$ is then given by $a_s=-\beta/\alpha$ \cite{AnalyticsFlambaum1}}. In Fig.\,\ref{fgr:ScatteringWavefunctions} we show the \textit{s}-wave scattering wavefunctions $\psi_0(r)$ for a collision energy $E_{coll}=k_B \times 1.0\, \mu\mathrm{K}$, with $k_B$ the Boltzmann constant, where the effect of the scattering length on both the asymptotic phase shift and the short-range scattering amplitude is evident. The \textit{s}-wave scattering lengths derived from the fitted model for $^{87}\mathrm{Rb}^{84}\mathrm{Sr}$ and $^{87}\mathrm{Rb}^{88}\mathrm{Sr}$ are in good agreement with those extracted from the cross-thermalization measurements presented in Sec.\,\ref{sec:CrossThermalizationMeasurements}, which corroborates the overall analysis carried out to this point.

\begin{table}[ht]
\small
\caption{Inter-species \textit{s}-wave scattering lengths in units of the Bohr radius}
\begin{tabular*}{\columnwidth}{@{\extracolsep{\fill}}lllll}
\hline
\noalign{\smallskip}
 & $^{84}\mathrm{Sr}$ & $^{86}\mathrm{Sr}$ & $^{87}\mathrm{Sr}$ & $^{88}\mathrm{Sr}$ \\
\hline
\noalign{\smallskip}
$^{85}\mathrm{Rb}$ & $689(20)$ & $90.6(2)$ & $44.3(3)$ & $-35.8(1.0)$\\
$^{87}\mathrm{Rb}$ & $92.7(2)$ & $-43.0(1.1)$ & $1421(98)$ & $170.3(6)$\\
\hline
\label{tbl:SwaveScatteringLengths}
\end{tabular*}
\end{table}

\subsubsection{Validation and inclusion of Fano-Feshbach spectroscopy.}
\label{sec:inclusionFFR}
The model described in the previous section is also \textit{sufficient} to infer within a few Gauss the resonant magnetic field of the magnetically-tunable Fano-Feshbach resonances (FFRs) present in RbSr. Let us note that we can only derive FFR locations from the fitted PEC with experimental accuracy thanks to the extreme simplicity of the $^2\Sigma^+$ ground state of RbSr \cite{AKAKE-I,RbSrFFR}. The existence and observability of this novel type of FFRs was theoretically predicted a few years ago \cite{PiotrRbSrMagnetoassociation} and recently experimentally observed by some of the authors \cite{RbSrFFR}.

The best fit $V_{LJ}$ in the previous section (see eqn\,(\ref{Lennard_Jones})) predicts the location of FFRs for fermionic $^{87}$Rb$^{87}$Sr within $10\,\mathrm{G}$ and has been used to infer with the same accuracy the location of one $^{87}\mathrm{Rb}^{88}\mathrm{Sr}$ FFR arising from the level $\{ \nu=64,\,N=0,\,F=2 \}$, subsequently observed in an ultracold Rb-Sr mixture \cite{RbSrFFR}. The BEs and corresponding quantum numbers of the bound levels inducing the observed FFRs derived with our model are reported in Table\,\ref{tbl:WeaklyBoundSpectrumLJ}, and marked with ``FFR'' in the ``Method'' column. As a complementary check, we apply the fitting procedure to the data set including both PA and FFRs, which results in the \textit{same} \textit{unique} solution. As in the case of two-colour PA spectroscopy data alone, inclusion of $C_6$ and $C_8$ is \textit{necessary} and \textit{sufficient} to model the complete data set. The best fit parameters are $\sigma=5.02477864619132\,$\r{A}, $C_6 = 1.776513404206001 \times 10^7\,$\r{A}$^6\,\mathrm{cm}^{-1}$, $C_8 = 6.262096495696839 \times 10^8\,$\r{A}$^8\,\mathrm{cm}^{-1}$, with $\mathrm{DOF}=8$ and $\widetilde{\chi}_{\mathrm{min}}^2=1.29$. The fit returns the physical quantities $C_6 = 1.777(18) \times 10^7\,$\r{A}$^6\,\mathrm{cm}^{-1}$, $C_8 = 6.3(1.3) \times 10^8\,$\r{A}$^8\,\mathrm{cm}^{-1}$, and a corresponding $\Phi^0_{WKB}=67.4370(13)$. There is a significant increase in our figure of merit compared to Sec.\,\ref{sec:PrecisionFitPA}, which we attribute primarily to the inclusion of deeper $F=2$ levels with rather large $\delta E_{\mathrm{hf}}$, and secondarily to the change in $\mathrm{DOF}$. The inferred \textit{s}-wave scattering lengths are consistent with those presented in Table\,\ref{fgr:ScatteringWavefunctions}.

\begin{figure}[ht]
\centering
  \includegraphics[width=\columnwidth]{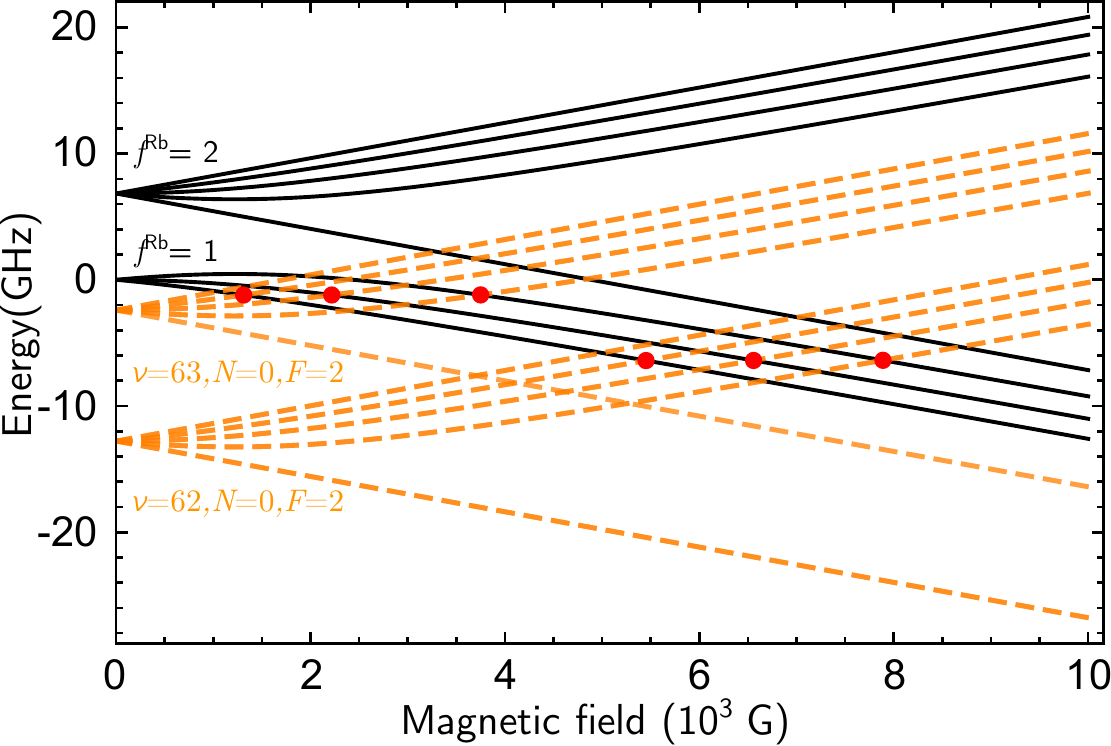}
  \caption{Zeeman sub-levels of $^{87}\mathrm{Rb}^{84}\mathrm{Sr}$. The plot shows the energy of $f^{\mathrm{Rb}}=1,2$ atomic levels (black solid lines) and of $F=2$ molecular levels (orange dashed lines, see also Table\,\ref{table:Feshbach Resonances}) in dependence of magnetic field. The red dots mark the location of FFRs in this magnetic field range.}
  \label{fgr:ZeemanLines}
\end{figure}

The ability to predict FFRs with high accuracy is extremely valuable for mixtures with one open-shell and one closed-shell atom, due to the low density of resonances in these systems, in particular in the case of zero nuclear magnetic moment for the  closed-shell atom, as in bosonic RbSr isotopologues \cite{PiotrRbSrMagnetoassociation, Brue2012LiYbFFR, Brue2013PRAFFR}. As an example of the outcomes of our model, Fig.\,\ref{fgr:ZeemanLines} shows the energy of the atomic scattering levels and molecular levels of $^{87}\mathrm{Rb}^{84}\mathrm{Sr}$ in dependence of magnetic field, and the locations of the predicted FFRs. Due to favourable scattering properties, this isotopic combination is a very good candidate for magneto-association \cite{BayerleThesis}.

\subsection{An independent check of quantum number assignment: inter-species thermalization}
\label{sec:CrossThermalizationMeasurements}
An improper quantum number attribution strongly affects the \textit{accuracy} of the inferred scattering lengths. We therefore experimentally characterize the thermalization of Rb atoms with a Sr cloud to measure the $\mathrm{Rb}\operatorname{-}\mathrm{Sr}$ inter-species \textit{s}-wave scattering lengths, and thus confirm our quantum number attribution. The values of scattering lengths obtained by thermalization experiments suffer from low \textit{precision}, however they constitute a useful cross-check, as they rely on simple collisional physics and are independent from our PA and Fano-Feshbach spectroscopy experiments. We present thermalization experiments done with $^{87}\mathrm{Rb}$-$^{84}\mathrm{Sr}$ and $^{87}\mathrm{Rb}$-$^{88}\mathrm{Sr}$. Trapped ultracold mixtures of $^{87}\mathrm{Rb}$-$^{87}\mathrm{Sr}$ show significantly stronger 3-body losses, which limit the reliability of the data analysis for this particular mixture.

\subsubsection{Experimental setup and sample conditions.}
\label{sec:CrossThermalizationDescription}
The starting point of the thermalization measurement is an ultracold mixture composed of $^{87}\mathrm{Rb}$ and either $^{84}\mathrm{Sr}$ or $^{88}\mathrm{Sr}$, prepared as for spectroscopy experiments, with the addition of evaporative cooling by lowering the dipole trap potential in $6\,\mathrm{s}$, followed by a $1\,\mathrm{s}$ re-compression of the potential, which is used to tune the atomic density and temperature. After this preparation sequence, the sample has a temperature between $200$ and $400\,\mathrm{nK}$ and contains $1-1.7 \times 10^5$ atoms of $^{84,88}\mathrm{Sr}$ and $50-70 \times 10^3$ atoms of $^{87}\mathrm{Rb}$. The typical shot-to-shot temperature fluctuation is $15\,\mathrm{nK}$, while shot-to-shot atom number fluctuations are $15\times 10^3$ and $5\times 10^3$ for Sr and Rb, respectively. The trapping frequencies in our crossed-beam dipole trap are $\{\omega_x, \,\omega_y,\, \omega_z \}=2 \pi \times \{35-65,\, 20-55, \, 500(25) \}\,\mathrm{Hz}$ for Sr and $\{\omega_x,\, \omega_y,\, \omega_z \}=2 \pi \times \{ 60-105,\, 35-90,\, 840(40) \}\,\mathrm{Hz}$ for Rb, respectively. The frequency ranges in the horizontal $x$ and $y$ axes correspond to various trap re-compressions. The relative uncertainty on these frequencies is less than $5\,\%$. The difference of trap frequencies between Sr isotopes is less than the uncertainty, hence negligible. The atomic densities used here are $n_{\mathrm{Sr}}=0.3 - 3 \times 10^{12} \, \mathrm{cm}^{-3}$, $n_{\mathrm{Rb}}= 0.7 - 4.4 \times 10^{12}\, \mathrm{cm}^{-3}$ for the $^{87}\mathrm{Rb}$-$^{84}\mathrm{Sr}$ mixture and $n_{\mathrm{Sr}}=1.7 - 4.4 \times 10^{12}\, \mathrm{cm}^{-3}$, $n_{\mathrm{Rb}}=1.5 - 5.4\times 10^{12}\, \mathrm{cm}^{-3}$ for the $^{87}\mathrm{Rb}$-$^{88}\mathrm{Sr}$ mixture. The Rb sample, as in PA spectroscopy, is prepared in $f^{\mathrm{Rb}}=1$ and is not spin-polarized. Given the existence of a single electronic ground state, $a_s$ can be considered independent of $f^{\mathrm{Rb}}$ and $m_f^{\mathrm{Rb}}$.

\subsubsection{Measurement strategy.}
\label{sec:Measurement strategy}
In order to observe inter-species collisions, we selectively excite the cloud of one species and observe the ensuing inter-species thermalization. Since the dipole trap is roughly three times deeper for Rb than for Sr, we excite the Rb cloud by scattering photons on Rb $\mathrm{D}_2$ line for a few $\mu\mathrm{s}$. After this excitation, the mixture is kept in the trap for a variable hold time $t$ before a $17\,\mathrm{ms}$ time-of-flight expansion followed by absorption imaging. From the absorption images, we extract temperatures and atom numbers of both species. The main limitations to the precision of our measurement are shot-to-shot fluctuations in atom number and temperature.

\subsubsection{Experimental results.}
\label{sec:CrossThermalizationResults}
\begin{figure*}[ht]
\includegraphics[width=\textwidth]{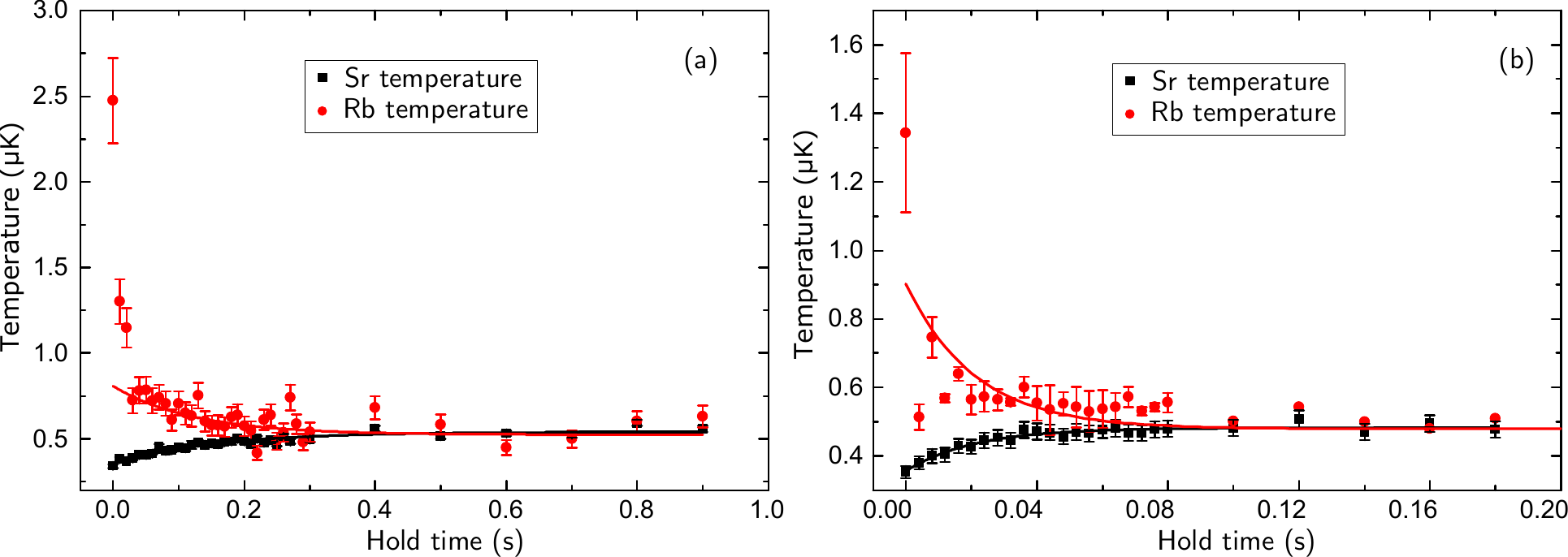}
\caption{Inter-species thermalization. Evolution of Sr (black squares) and Rb (red circles) temperature as a function of the hold time during thermalization in (a) $^{87}\mathrm{Rb}\operatorname{-}^{84}\mathrm{Sr}$ mixture at an effective flux of $\Phi = 7.0\times 10^{12}\,\mathrm{s}^{-1}\,\mathrm{cm}^{-2}$ and (b) $^{87}\mathrm{Rb}\operatorname{-}^{88}\mathrm{Sr}$ mixture at a flux of $\Phi = 1.9\times 10^{13}\,\mathrm{s}^{-1}\,\mathrm{cm}^{-2}$. The lines are exponential fits to the data.}
\label{fig:ExampleCrossThermalization}
\end{figure*}

We measure the evolution of temperature and atom number for each species as functions of time. In Fig.\,\ref{fig:ExampleCrossThermalization} we show an example for each isotopic combination. The temperature of Sr smoothly evolves from the initial temperature $T_E^i=T_{\mathrm{Sr}}(t=0)$ to the final equilibrium temperature $T_E^f=T_{\mathrm{Sr}}(t\rightarrow\infty)$. By contrast, the temperature of Rb shows a sharp decrease on a timescale of a few tens of ms from a temperature of a few $\mu\mathrm{K}$ down to $0.5-0.7\,\mu\mathrm{K}$, after which the new equilibrium temperature $T_E^f=T_{\mathrm{Rb}}(t\rightarrow\infty)=T_{\mathrm{Sr}}(t\rightarrow\infty)$ is reached smoothly.

\subsubsection{Extraction of collision cross sections.}
\label{sec:Data Analysis and Conclusions}
\begin{figure}[ht]
\includegraphics[width=\columnwidth]{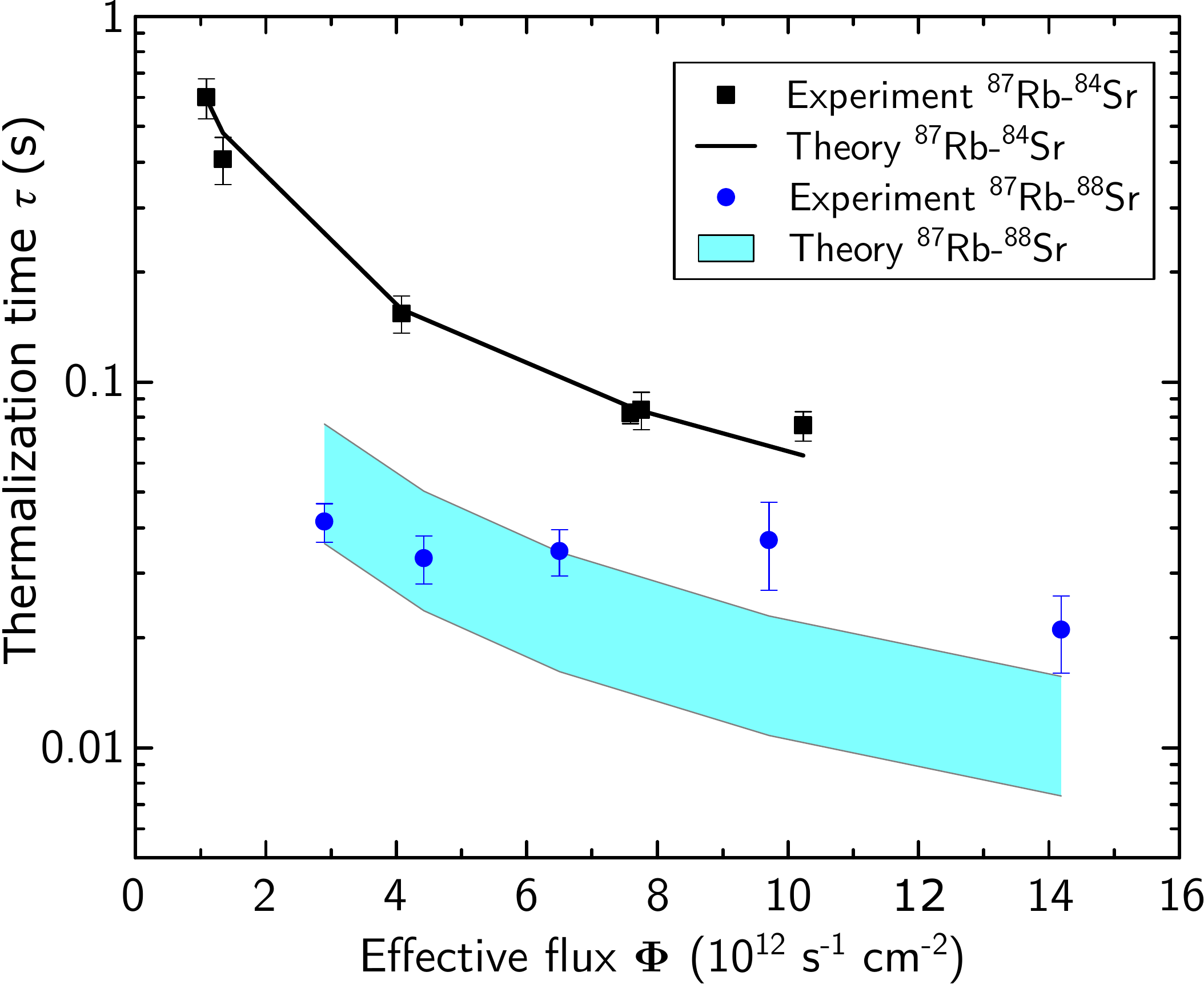}
\caption{Thermalization time as function of the effective flux $\Phi$ (see main text) for both isotopic combinations. Black squares are data for $^{87}\mathrm{Rb}\operatorname{-}^{84}\mathrm{Sr}$ and blue circles for $^{87}\mathrm{Rb}\operatorname{-}^{88}\mathrm{Sr}$. The solid black line shows the fit for $^{87}\mathrm{Rb}\operatorname{-}^{84}\mathrm{Sr}$ using the theory described in the text. The light-blue area shows the theoretical region corresponding to the uncertainty of the fitted scattering length for $^{87}\mathrm{Rb}\operatorname{-}^{88}\mathrm{Sr}$. }
\label{fig:CrossThermalizationFit}
\end{figure}

We measure the thermalization time of both $^{87}\mathrm{Rb}\operatorname{-}^{84}\mathrm{Sr}$ and $^{87}\mathrm{Rb}\operatorname{-}^{88}\mathrm{Sr}$ mixtures. In the case of close-to-equilibrium dynamics, the evolution of temperatures $T_{\mathrm{Rb},\mathrm{Sr}}(t)$ is described by exponential functions with the same well-defined time constant $\tau$. We analyse the thermalization rate $\tau^{-1}$ using a well-known model \cite{MoskCrossThermalization,DalibardColdCollisions}, which we detail in Appendix\,\ref{sec:CrossThermalizationTheory}. This model gives the relation:

\begin{equation}
\label{SimplifiedMainTextCrossThermRate}
\tau^{-1}\approx\frac{1}{2.4} \, \sigma_{\mathrm{Rb}\operatorname{-}\mathrm{Sr}} \times \Phi,
\end{equation}

\noindent
where $\sigma_{\mathrm{Rb}\operatorname{-}\mathrm{Sr}} = 4 \pi \, a^2_{\mathrm{Rb}\operatorname{-}\mathrm{Sr}}$ is the collision cross section dependent on the inter-species \textit{s}-wave scattering length $a_{\mathrm{Rb}\operatorname{-}\mathrm{Sr}}$. The value 2.4 in the denominator represents the average number of collisions required for thermalization, when thermalization is fast compared to the trap frequencies. $\Phi$ is an effective flux that encompasses the kinematic contribution, see Appendix\,\ref{sec:CrossThermalizationTheory}. We fit our data for various effective fluxes and extract values for $\tau$, shown in Fig.\,\ref{fig:CrossThermalizationFit}.

The effective flux $\Phi$ is determined through the knowledge of the trap potential, atom numbers and initial temperatures. All quantities are either measured or known from calibration, with the exception of the initial temperature $T_{\mathrm{Rb}}^i$ of the Rb sample right after excitation. Let us note that the excitation we apply experimentally is the injection of energy in the form of both heating and displacement of the cloud. However, by assuming the regime of close-to-equilibrium dynamics, we approximate the excitation to be solely an increase in temperature. The excitation energy of Rb can be derived with good precision from the atom numbers and the temperature evolution of Sr, since the system is isolated after the excitation. The trapping potential can be approximated at these low temperatures by a three dimensional harmonic oscillator potential giving $E=3 \, k_B T$ energy per particle. The final energy in the system $E^f$ must be equal to the initial one $E^i$, and under our assumptions these are $E^f=3 \,k_B T_E^f (N_{\mathrm{Sr}}+N_{\mathrm{Rb}})$ and $E^i=3 \,k_B (N_{\mathrm{Sr}}T_E^i+ N_{\mathrm{Rb}}T_{\mathrm{Rb}}^i)$. We thus derive $T_{\mathrm{Rb}}^i =T_E^f + \frac{N_{\mathrm{Sr}}}{N_{\mathrm{Rb}}} (T_E^f - T_E^i)$. 

Fitting the data of Fig.\,\ref{fig:CrossThermalizationFit} with eqn\,(\ref{SimplifiedMainTextCrossThermRate}), we obtain the inter-species scattering lengths $|a_{^{87}\mathrm{Rb}\operatorname{-}^{84}\mathrm{Sr}}|=103_{-10}^{+15} \,a_0$ and $|a_{^{87}\mathrm{Rb}\operatorname{-}^{88}\mathrm{Sr}}|=215_{-40}^{+50} \,a_0$, where $a_0$ is the Bohr radius, and where the errors are estimated from the residual sum of squares 5-fold increase. Let us note that the variation of the average number of collisions required for thermalization, within the meaningful range $2.4-3.0$ \cite{DalibardColdCollisions}, leads to a variation of the scattering lengths smaller than the stated error.

While for $^{87}\mathrm{Rb}\operatorname{-}^{84}\mathrm{Sr}$ the fit is satisfying, the fit of $^{87}\mathrm{Rb}\operatorname{-}^{88}\mathrm{Sr}$ is worse because of the two points at highest $\Phi$, which we include in the fit. For these two points the thermalization time is comparable with the initial fast time scale of the Rb temperature evolution (see Fig.\,\ref{fig:ExampleCrossThermalization}), suggesting a strong deviation from the close-to-equilibrium case, as expected from the bigger inter-species scattering length. Nonetheless even in the latter case, a meaningful scattering length can be extracted with a correspondingly (larger) error. Finally, the $^{87}\mathrm{Rb}\operatorname{-}^{87}\mathrm{Sr}$ mixture shows losses that we interpret as 3-body losses, which for similar densities are not observed in the other mixtures investigated. From this observation, we derive that $|a_{^{87}\mathrm{Rb}\operatorname{-}^{87}\mathrm{Sr}}| \gg |a_{^{87}\mathrm{Rb}\operatorname{-}^{88}\mathrm{Sr}}|\simeq 200 \, a_0$.

The fitted (central) values of the inter-species scattering lengths are close to the ones inferred from spectroscopy, which is an independent confirmation of our quantum number attribution. However they are $10\,\% - 20\,\%$ higher. This is expected since the initial densities of Rb are underestimated by our model, which assumes thermalization. A Monte-Carlo trajectory simulation would most likely improve the accuracy of the scattering lengths extracted from these thermalization measurements.

\section{Thermoluminescence and LIF spectroscopy of deeply-bound levels}
\label{sec:HotExpPart}

In the second experiment, we study the deeply-bound levels supported by the $\mathrm{B}(2)^{2}\Sigma^{+}$ and $\mathrm{X}(1)^{2}\Sigma^{+}$ potentials via fluorescence spectroscopy of a $1000 \,\mathrm{K}$ gas mixture of Rb and Sr. In this section, we first describe the experimental setup we use to record the fluorescence from RbSr molecules in a heat-pipe oven. We then explain how we simulate theoretical spectra using three published sets of potential energy curves produced by independent \textit{ab-initio} methods \cite{PiotrOlivierRbSr, ErnstLatestPECs}. By comparing these spectra with our experimental data, we identify a few band heads, from which we extract Dunham coefficients describing the deepest parts of the $\mathrm{B}(2)^{2}\Sigma^{+}$ and $\mathrm{X}(1)^{2}\Sigma^{+}$ states. With the obtained two sets of Dunham coefficients, we repeat the comparison procedure until we identify 24 band heads and produce final sets of Dunham coefficients. Finally, we estimate the uncertainty of the Dunham coefficients resulting from our analysis by a Monte-Carlo method.

\subsection{Experimental setup}
\label{subsec:HotExpPart.Exp}
The measurements at high temperatures were performed in two steps. In the first step, we record thermoluminescence spectra using the method and experimental setup described in ref.~\citenum{JacekKSR}. We therefore provide here only information specific to this paper. We produce RbSr molecules in a dedicated dual-temperature heat-pipe oven. We place $10\,\mathrm{g}$ of metallic strontium in the central part of the oven, which is heated to $T_{\mathrm{Sr}} = 1000 \,\mathrm{K}$, and $8\,\mathrm{g}$ of metallic rubidium in the outer part, heated to $T_{\mathrm{Rb}} = 800 \,\mathrm{K}$. Both metals have natural isotopic composition. To ensure the stability of the heat-pipe operation, we use a buffer gas of helium at a pressure of $30\,\mathrm{Torr}$. At the applied temperatures the $\mathrm{B}(2)^{2}\Sigma^{+}$ electronic state of RbSr is thermally populated, and we record the fluorescence towards the $\mathrm{X}(1)^{2}\Sigma^{+}$ electronic ground state using a Bruker Vertex V80 Fourier Transform Spectrometer with a spectral resolution of $0.16\,\mathrm{cm}^{-1}$ limited by its aperture size.

In the second step, we obtain spectra via laser induced fluorescence (LIF). We employ a home-made $100\,\mathrm{mW}$ external-cavity diode laser whose wavelength is actively stabilized using a HighFinesse WS7 wavemeter. By tuning the laser frequency to the centre of selected band heads, we excite RbSr molecules to the $\mathrm{B}(2)^{2}\Sigma^{+}$ state and record fluorescence to the ground state with the same spectrometer as before. To increase the contrast between the LIF and thermoluminescence signals observed simultaneously, we reduce the temperature of the central part of the heat-pipe to $T_{\mathrm{Sr}} = 900\,\mathrm{K}$.

\subsection{Simulations of the recorded spectra}
\label{subsec:HotExpPart.SpectraSimulation}

In order to interpret the experimental spectra, we first simulate fluorescence spectra using PECs and transition dipole moments computed theoretically, and compare theory and experiment. The simulations start from three sets of PECs, obtained independently with FCI-ECP+CPP \cite{PiotrOlivierRbSr}, RCCSD(T) \cite{PiotrOlivierRbSr}, and MRCI \cite{ErnstLatestPECs} methods. We calculate the energies of rovibrational levels of the $\mathrm{B}(2)^{2}\Sigma^{+}$ and $\mathrm{X}(1)^{2}\Sigma^{+}$ states by solving the radial Schr\"{o}dinger equation with each of the three sets of PECs. All bound levels in the $\mathrm{X}(1)^{2}\Sigma^{+}$ and $\mathrm{B}(2)^{2}\Sigma^{+}$ states are included in the simulations. The contribution of the $\mathrm{A}(1)^{2}\Pi$ state in this spectral region was found to be negligible. In our calculation we omit the fine structure splitting of molecular levels resulting from spin-rotational coupling. Indeed, the energy difference between fine structure components with low rotational quantum numbers $N'$ contributing to a band head formation is expected to be smaller than the spectral resolution of the measurement \cite{Tiemann2017KCa}. We assume spectral lines to have a Gaussian profile with $\mathrm{FWHM}=0.16\,\mathrm{cm}^{-1}$, which results from the Fourier Transform Spectrometer working parameters. Intensities of all spectral lines are calculated assuming thermal equilibrium in the central part of the heat-pipe. The simulation procedure has been described in detail by \citet{JacekKSR}, including equations necessary to perform the calculations.

\begin{figure}[ht]
\centering
  \includegraphics[width=0.95 \columnwidth]{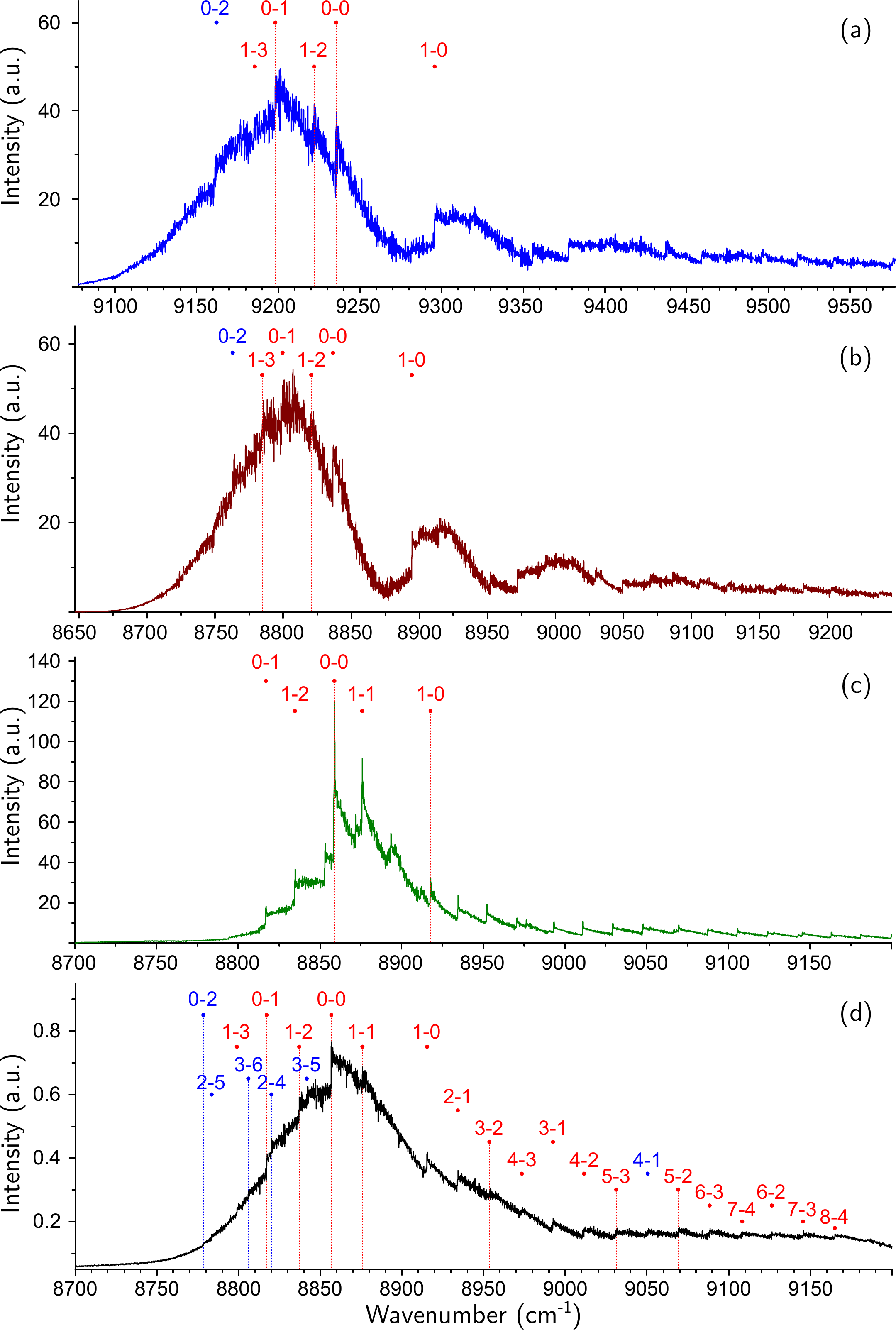}
  \caption{Simulated thermoluminescence spectra based on the three sets of theoretical PECs for the $\mathrm{X}(1)^{2}\Sigma^{+}$ and $\mathrm{B}(2)^{2}\Sigma^{+}$ states, calculated with (a) RCCSD \cite{PiotrOlivierRbSr}, (b) FCI-ECP+CPP \cite{PiotrOlivierRbSr}, and (c) MRCI \cite{ErnstLatestPECs}, compared with the experimental data (d). The positions of identified band heads are marked with dashed lines and labelled with their vibrational quantum numbers $\nu'-\nu''$ (where primed and double primed symbols refer to $\mathrm{B}(2)^{2}\Sigma^{+}$ and $\mathrm{X}(1)^{2}\Sigma^{+}$, respectively). The wavenumber scales of theoretical spectra are adjusted in such a way that $0-0$ band heads are at the same position in all panels. The well-resolved band heads used in the final fit of Dunham coefficients are marked in red.}
  \label{fgr:RbSr_spectrum}
\end{figure}

The final step of the calculations is to average the simulated spectra of the most abundant isotopologues of RbSr, weighted by their natural abundances ($59.6\,\%$ for $^{85}\mathrm{Rb}^{88}\mathrm{Sr}$, $22.9\,\%$ for $^{87}\mathrm{Rb}^{88}\mathrm{Sr}$, $7.1\,\%$ for $^{85}\mathrm{Rb}^{86}\mathrm{Sr}$, $5.1\,\%$ for $^{85}\mathrm{Rb}^{87}\mathrm{Sr}$, $2.7\,\%$ for $^{87}\mathrm{Rb}^{86}\mathrm{Sr}$ and $1.9\,\%$ for $^{87}\mathrm{Rb}^{87}\mathrm{Sr}$). As a result, we obtain three sets of ``theoretical spectra'' to be compared with the experimental data, shown in Fig.\,\ref{fgr:RbSr_spectrum}. The analysis of the spectra reveals that the positions of the observed band heads are defined by the $^{85}\mathrm{Rb}^{88}\mathrm{Sr}$ isotope alone, and other isotopes influence mainly the band-head widths (broadened up to $0.08\,\mathrm{cm}^{-1}$). Thus we only take into account the $^{85}\mathrm{Rb}^{88}\mathrm{Sr}$ isotope in the Dunham coefficients generation procedure described in the next subsection. The influence of other isotopes is included again during the error estimation process.

\subsection{Results}
\label{subsec:HotExpPart.Results}
In order to identify the observed band heads, we compare the thermoluminescence spectra of RbSr with the simulated spectra based on the three theoretical methods \cite{PiotrOlivierRbSr, ErnstLatestPECs}. Unfortunately, these simulations provide spectra of considerably different shapes for each theory (see Fig.\,\ref{fgr:RbSr_spectrum}) and only few experimental band heads can be identified unambiguously as they appear in all three simulations.

\begin{figure}[ht]
\centering
\includegraphics[width=0.95 \columnwidth]{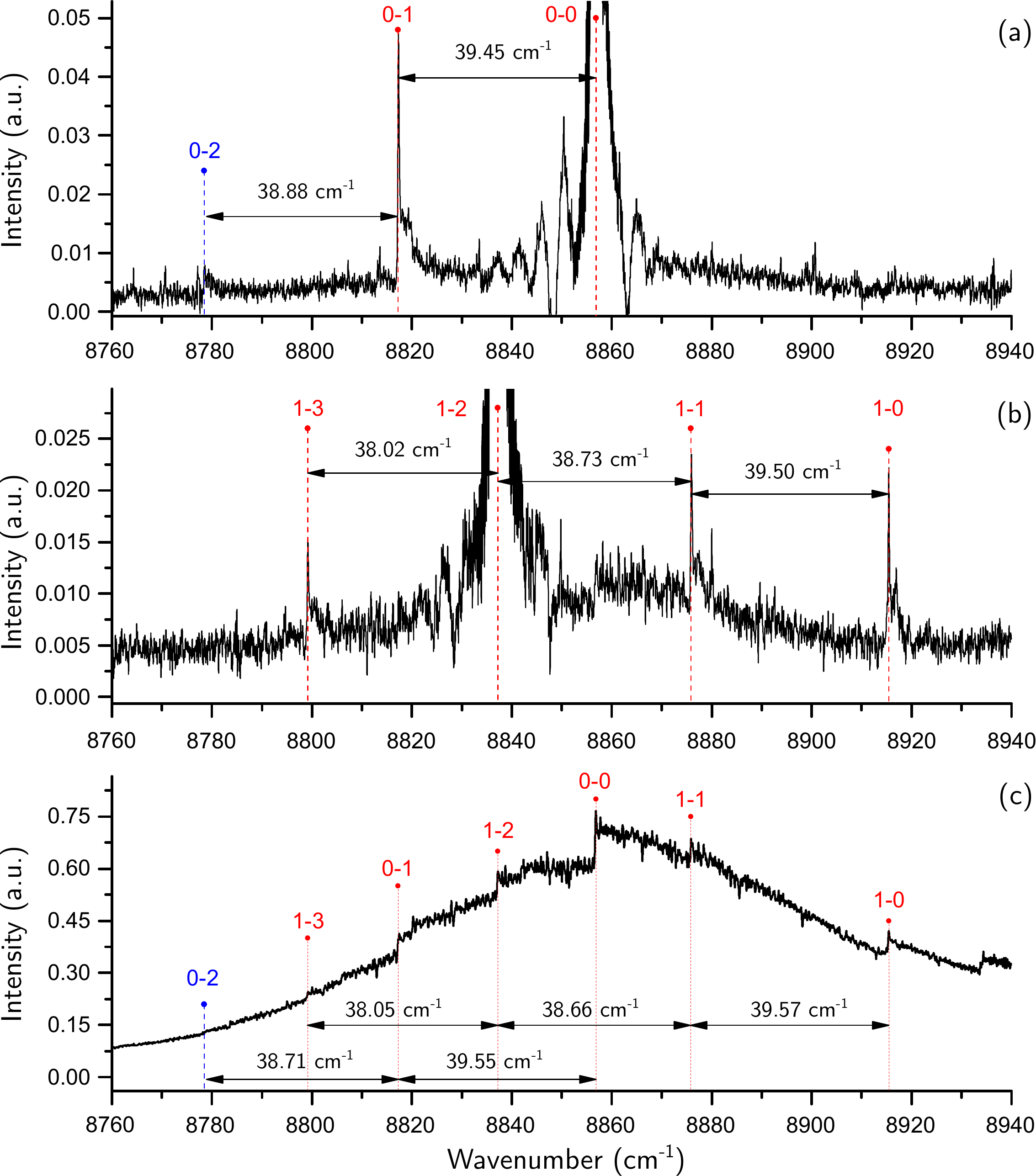}
\caption{Comparison of band-head positions in the LIF (a) and (b), and thermoluminescence (c) spectra. The LIF spectra were obtained with the laser tuned to the centres of $\nu'-\nu''$ band heads: (a) $0-0$ and (b) $1-2$. The well-resolved band heads used in the preliminary fit of Dunham coefficients are marked in red.
}
\label{fgr:RbSr_LIF}
\end{figure}

To address this issue we record additional LIF spectra by tuning the excitation laser frequency to the centres of already identified band heads. These new experimental data, shown in Fig.\,\ref{fgr:RbSr_LIF}, confirm the validity of the assignment in the case of six band heads. Using the energy of experimental band heads and their assignment confirmed both by thermoluminescence and LIF spectroscopy, we fit preliminary Dunham coefficients for both $\mathrm{X}(1)^{2}\Sigma^{+}$ and $\mathrm{B}(2)^{2}\Sigma^{+}$ electronic states. The values of the ground state rotational constants (labelled $Y_{01} \equiv B_e$) were taken from theory for each set and fixed during the fit. We thus obtain three sets of fitted coefficients, each corresponding to one theoretical method. This procedure is described in detail in ref.~\citenum{JacekKSR}.

These fitted Dunham coefficients allow for a new prediction of the vibrational level energies in the $\mathrm{X}(1)^{2}\Sigma^{+}$ and $\mathrm{B}(2)^{2}\Sigma^{+}$ states, followed by an assignment of additional band heads in the thermoluminescence spectrum. With the improved assignment a correction of Dunham coefficients becomes possible, and we repeat the whole procedure until the final identification of 24 band heads, whose energies are given in Table\,\ref{tab:deslandres_exp_RbSr} as a Deslandres table. To prevent mistakes in the assignment, only the 18 strongest band heads, whose energies are written in bold in the table and whose positions are marked in red in Fig.\,\ref{fgr:RbSr_spectrum}, were taken into account in the final fit of Dunham coefficients. As the outcome of this hot gas mixture spectroscopy, the fitted coefficients describe the energies of the six lowest vibrational levels $\nu^{\prime\prime} = 0-5$ in the ground state and the nine lowest vibrational levels $\nu^{\prime} = 0-8$ of the $\mathrm{B}(2)^{2}\Sigma^{+}$ state. The final values of the Dunham coefficients, given in Table\,\ref{tab:RbSrDunham}, will be used in the next steps of our analysis described in the following section.

\begin{table*}[!ht] 
	\caption{Deslandres table constructed for the observed band heads in the experimental thermoluminescence spectrum of RbSr. The wavenumbers of band heads are given in $\mathrm{cm}^{-1}$. The energies of the 18 strongest band heads used in the final fit are written in bold}
	\label{tab:deslandres_exp_RbSr}
	\begin{tabular*}{\textwidth}{@{\extracolsep{\fill}}llllllllllllll}
		
		\hline	      & $\nu^{\prime\prime}=0$    &         & $1$ &     & $2$ &   & $3$ & & $4$ & &$5$&  & $6$ \\ \hline
		$\nu^{\prime}=0$ & $\textnormal{\textbf{8856.81}}$ & $39.55$ & $\textnormal{\textbf{8817.26}}$ &38.71   &  8778.55 & &   &   & & &  && \\ 
		& $58.63$   &         & $58.61$ &   &$58.66$   &   &   &   & & & &&\\ 
		$1$ & $\textnormal{\textbf{8915.44}}$ & $39.57$ & $\textnormal{\textbf{8875.87}}$ & $38.66$ & $\textnormal{\textbf{8837.21}}$ & $38.05$&$\textnormal{\textbf{8799.16}}$ & & & & &&  \\ 
		&           &         & $58.52$ &  &  &  &  &  & & & & &\\
		$2$ &           &         & $\textnormal{\textbf{8934.39}}$ &   &   &   &    & &$8820.25$ & 36.6 & 8783.65& & \\ 
		&           &         & $58.18$ &   &   &   &   & & & & $58.2$& & \\ 
		$3$ &           &         & $\textnormal{\textbf{8992.57}}$ & $38.89$ & $\textnormal{\textbf{8953.68}}$ &&  & & & &$8841.85$&$ 35.84$&$8806.01$ \\
		&           &         & $58.06$ &           & $57.92$ & &  & & & & & & \\ 
		$4$ &           &         & $9050.63$ & $39.03$ & $\textnormal{\textbf{9011.60}}$ &$37.97$  &$\textnormal{\textbf{8973.63}}$ & & & & & & \\ 
		&           &         &           &         & $57.6$ & &$57.53$  & & & & & &\\ 
		$5$ &  &  &  &  & $\textnormal{\textbf{9069.2}}$ & $38.04$ & $\textnormal{\textbf{9031.16}}$ &  & & & & &\\ 
		&     &     &     &     & $57.34$   &    & $57.35$ &  & & & & &  \\ 
		$6$ &     &     &     &     &   $\textnormal{\textbf{9126.54}}$ &  $38.03$  & $\textnormal{\textbf{9088.51}}$ & & & & & & \\ 
        &     &     &     &     &   &    & $57.17$  &  & & & & &  \\ 
		$7$ &   &   &   &   &   &   &  $\textnormal{\textbf{9145.68}}$    &$37.33 $  &$\textnormal{\textbf{9108.35}} $ & & & & \\ 
        &     &     &     &     & &    &  &  &$56.82$ & & & &  \\ 
        $8$ &   &   &   &   &   &   &      & &$\textnormal{\textbf{9165.17}}$ & & & &  \\ \hline
	\end{tabular*}
\end{table*}

\begin{table}[!ht]
	\caption{The Dunham coefficients for the $\mathrm{B}(2)^{2}\Sigma^{+}$ and $\mathrm{X}(1)^{2}\Sigma^{+}$ states of the $^{85}\mathrm{Rb}^{88}\mathrm{Sr}$ molecule based on the LIF and thermoluminescence spectra. The three sets of coefficients for each state correspond to different values of the ground state equilibrium distance $r_e$ taken from theoretical calculations. All values are in $\mathrm{cm}^{-1}$}
	\label{tab:RbSrDunham}
	\begin{tabular}{llll}	
		\hline
        \noalign{\smallskip}
		& MRCI \cite{ErnstLatestPECs} & FCI-ECP+CPP \cite{PiotrOlivierRbSr}& RCCSD(T) \cite{PiotrOlivierRbSr}\\ \hline
        \noalign{\smallskip}
		&\multicolumn{3}{c}{$\mathrm{X}(1)^{2}\Sigma^{+}$}\\
		\multicolumn{1}{c}{$Y_{10}$} & $40.39(72)$ & $40.32(76)$& $40.31(76)$\\
		\multicolumn{1}{c}{$Y_{20}$ }& $-0.39(11)$& $-0.38(12)$& $-0.38(12)$ \\
		\multicolumn{1}{c}{$Y_{01}\times 10^{2}$} & $1.874^{a}$&$1.84842^{a}$&$1.79052^{a}$\\
		\multicolumn{1}{c}{$Y_{11}\times 10^{4}$} & $-0.8(6)$& $-1.1(1.1)$ & $-0.9(1.2)$\\\hline
        \noalign{\smallskip}
		\multicolumn{1}{c}{}&\multicolumn{3}{c}{$\mathrm{B}(2)^{2}\Sigma^{+}$}\\
		\multicolumn{1}{c}{$T_e$} & $8847.92(80)$&$8847.66(80)$&$8847.66(80)$ \\
		\multicolumn{1}{c}{$Y_{10}$} & $58.96(38)$& $58.95(39)$& $58.95(39)$\\
		\multicolumn{1}{c}{$Y_{20}$ }& $-0.13(4)$ & $-0.13(5)$ & $-0.13(5)$\\
		\multicolumn{1}{c}{$Y_{01}\times 10^{2}$ }& $1.932(4)^{b}$& $1.952(5)^{b}$& $1.893(5)^{b}$\\
		\multicolumn{1}{c}{$Y_{11}\times 10^{5}$}  & $-8.3(6.4)$& $-3.4(6.7)$&$-3.4(6.8)$ \\ \hline
	\end{tabular}\\
	$^{a}$ Values taken from theory and fixed during the fit.\\
	$^{b}$ Values strongly correlated with $Y_{01}$ of the $\mathrm{X}(1)^{2}\Sigma^{+}$ state. \\
\end{table}

The uncertainties of the Dunham coefficients result mainly from the determination of the positions and widths of the band heads, as many lines corresponding to transitions between different rovibrational levels of the $\mathrm{B}(2)^{2}\Sigma^{+}$ and $\mathrm{X}(1)^{2}\Sigma^{+}$ states overlap in the spectra, and thus only the top parts of the band heads are observed in our experiment. We use a Monte-Carlo method to find the error associated with this problem. We randomly vary the positions of individual band heads within a range of $0.46\,\mathrm{cm}^{-1}$. The choice for this range results from the band-head half-widths, assumed arbitrarily to be $0.3\,\mathrm{cm}^{-1}$, combined with the maximum value of the isotopic shifts. We also vary the widths of individual band heads within $0.16\,\mathrm{cm}^{-1}$, a value that influences the number of rovibrational lines taken into consideration in each case. We determine a set of Dunham coefficients for each random combination of positions and widths. We repeat the procedure until the average values of all coefficients becomes equal to the fitted values reported in Table\,\ref{tab:RbSrDunham}. The final errors are defined for each Dunham coefficient as three times their standard deviation.

\section{\textit{Ab initio}-based PEC fit}
\label{sec:JointPart}

We now combine the results from both types of spectroscopy, at $\mu\mathrm{K}$ and $1000\,\mathrm{K}$ temperatures, and perform a joint fit procedure in order to obtain a model representing the complete spectrum of the $\mathrm{X}(1)^{2}\Sigma^{+}$ ground state of RbSr. In this section, we first describe the specifics of the problem we will address. We then submit and motivate our choice of representation for the potential energy curves. We next detail all steps of our fitting procedure. Finally, we present the results of our joint analysis and compare them with the predictions of the three \textit{ab-initio} theoretical methods we selected.

\subsection{Statement of the problem}
\label{subsec:JointPart.Problem}
The goal of our data analysis is to provide a representation of the complete bound spectrum of the electronic ground state. This apparently contradicts the fact that, based on the independent analyses of the weakly- and deeply-bound levels, only $15\,\%$ of the vibrational levels, corresponding to less than $25\,\%$ of the well depth, were observed. Moreover, two-colour photoassociation spectroscopy has very high accuracy and precision but only a few weakly-bound levels have been probed, while thermoluminescence spectroscopy explored a significantly bigger energy range but lacks rotational resolution and its precision is limited to $0.16\,\mathrm{cm}^{-1}$. To the knowledge of the authors, such a problem has not been addressed before and requires a novel method of analysis able to exploit all information present in the two data sets at our disposal.

An examination of the methods used in the previous sections shows how to overcome this problem. The weakly-bound spectrum was analysed, without the need for \textit{ab-initio} PECs, via a direct potential fit of an analytic PEC, with the sole requirement of a correct long-range behaviour, see Sec.\,\ref{subsec:ColdExpPart.DataAnalysis}. The deeply-bound spectrum was analysed by a fit of Dunham expansion coefficients to band heads whose rovibrational composition was determined by the simulated spectrum based on \textit{ab-initio} PECs, see Sec.\,\ref{subsec:HotExpPart.SpectraSimulation} and \ref{subsec:HotExpPart.Results}. Since the fitted Lennard-Jones PEC and the Dunham expansion have no predictive power beyond the corresponding regions of definition, the results of those analyses are valid separately but cannot be extrapolated to the region with missing data. However, a model originating from \textit{ab-initio} calculations, with an appropriate PEC for the $\mathrm{X}(1)^2\Sigma^+$ state and a correlated $\mathrm{B}(2)^2\Sigma^+$ state is expected to be a good representation of the complete data set, capable of predictive power for the $\mathrm{X}(1)^2\Sigma^+$ state, and easily refined in the future by inclusion of new data.

\subsection{Representation of the \texorpdfstring{$\mathrm{X}(1)^2\Sigma^+$}{X(1)2Sigma+} and \texorpdfstring{$\mathrm{B}(2)^2\Sigma^+$}{B(2)2Sigma+} state PECs}
\label{subsec:JointPart.Xstate}
We now choose a suitable representation of the RbSr $\mathrm{X}(1)^2\Sigma^+$ and $\mathrm{B}(2)^2\Sigma^+$ states. A somewhat similar problem, albeit considerably more intricate, arose in the case of the excited $1^3\Sigma^+_g$ state in the homonuclear $\mathrm{Li}_2$ molecule accessed via the $1^3\Sigma^+_g\rightarrow \mathrm{a}^3\Sigma^+_u$ system, see ref.~\citenum{Dattani2011JointAnalysis}. In this case, \citeauthor{Dattani2011JointAnalysis} were able to bridge a $5000\,\mathrm{cm}^{-1}$ gap in spectroscopy data, i.e. $70\,\%$ of the well depth, by performing direct potential fit of Morse/Long-Range (MLR) functions to a rovibrationally resolved high-precision spectrum \footnote[1]{Successively, the binding energies extrapolated in the gap region were experimentally confirmed within $1.5\,\mathrm{cm}^{-1}$ \cite{MadisonLi2}.}. The MLR function appears to be particularly suited to represent the RbSr ground state. Indeed, since the RbSr $\mathrm{X}(1)^2\Sigma^+$ state is a single isolated electronic state and RbSr is a heavy molecule, the MLR PEC can easily represent the long-range tail, Born-Oppenheimer breakdown effects are expected to be small \cite{JeremyBOB} \footnote[1]{Adiabatic corrections are similar to those in $\mathrm{Rb}_2$ \cite{LeRoyRb2, VerhaarRb2}.}, and damping functions are readily incorporated \cite{LeRoyDampingFunctionsMLR}. The $\mathrm{B}(2)^2\Sigma^+$ state, relevant for thermoluminescence spectroscopy, is experimentally probed only close to its bottom, far from other electronic states, so it can be explicitly included in the analysis. However, compared to the case of ref.~\citenum{Dattani2011JointAnalysis}, the precision of our thermoluminescence data is significantly lower and lacks rotational resolution. As a consequence, for reasons different from the case of ref.~\citenum{Dattani2011JointAnalysis}, the fit is non-trivial and a specific method must be devised.

The version of the MLR function used in this work is the same as in ref.~\citenum{Dattani2011JointAnalysis}:
\begin{equation}
\label{MLR_function}
V_{\mathrm{MLR}}(r) =D_e \left[ 1- \frac{u(r)}{u(r_e)} e^{-\beta \left( y_p(r),y_q(r) \right) \cdot y_p(r)} \right]^2,
\end{equation}
\begin{equation}
\label{u_MLR}
u(r) = \sum_{i=1}^{N_u} d_{m_i}(r)\cdot \frac{C_{m_i}}{r^{m_i}},
\end{equation}
\begin{equation}
\label{RadialVariable_MLR}
y_x(r) = \frac{r^x-r_e^x}{r^x+r_e^x},
\end{equation}
\begin{equation}
\label{ExponentRadialVariable_MLR}
\beta \left( y_p,y_q \right) =\ln\left(\frac{2D_e}{u(r_e)}\right)\cdot y_p(r)  + \left(1-y_p(r)\right)\cdot \sum_{i=0}^{N_\beta}\beta_i \left( y_q(r) \right)^i,
\end{equation}
where $D_e$ is the well depth, $r_e$ is the equilibrium distance, $u(r)$ is the function describing the long-range behaviour, $y_x(r)$ is an $x$-order effective radial variable and $\beta \left( y_p,y_q \right)$ is the exponent coefficient of the radial variable $y_p$ \footnote[2]{Compared to ref.~\citenum{Dattani2011JointAnalysis} the notation is simplified because we set $r_{ref}=r_e$, i.e. all effective radial variables are referenced to $r_e$.}. The functions $d_{m_i}(r)$, explicitly included in eqn\,(\ref{u_MLR}), are Douketis-type \cite{DouketisDampingFunctions} damping functions $D_m^s(r)$ with $s=-1$ and adapted to RbSr by scaling of the radial variable via atomic ionization potentials \cite{NIST} as explained in ref.~\citenum{Dattani2011JointAnalysis}. The $C_{m_i}$ coefficients in eqn\,(\ref{u_MLR}) are the $N_u$ lowest order dispersion coefficients.

In order to represent the theoretical PECs by MLR functions we choose a family of these functions and values of their parameters based on the available theoretical calculations. The family of the MLR functions is defined by the choice of $N_u$ in eqn\,(\ref{u_MLR}), $N_\beta$ in eqn\,(\ref{ExponentRadialVariable_MLR}) and $p,q$. We use perturbation-theory results for $C_6$, $C_8$ and $C_{10}$ available in the literature \cite{CICPDispersionCoeffMitroy} and set $N_u=3$. This choice implies $p>m_{last}-m_1=4$ and correspondingly $1<q<p$ \cite{Dattani2011JointAnalysis}, with a contribution to the asymptotic long-range tail of order $r^{-m_1-p}=r^{-6-p}$. We resolve this indefiniteness, together with the one of $N_\beta$, by fitting the MLR function to the three point-wise representations of \textit{ab-initio} PECs in the region $r \geq 3.0$\,\r{A}, using the unweighted $\widetilde{\chi}^2$ as figure of merit, with errors set to 1.0\,cm$^{-1}$. In all fitted cases we obtain $\widetilde{\chi}_{\mathrm{min}}^2\simeq 1.0$ with ``well-behaved'' PECs, i.e. with a single inflection point, already for $N_\beta=5$, and the best fits are obtained for low values of $p,q$. Hence, we eventually set $N_u=3$, $N_\beta=5$, $p=5$ and $q=2$, which we hold constant during later fits. Since our data on deeply-bound levels are not rotationally resolved and since weakly-bound levels, within our experimental precision, do not carry information on the equilibrium distance $r_e$, we set $r_e$ equal to the equilibrium distances from the theoretical calculations and hold it fixed during fits. The remaining parameters $D_e$ and $\beta_i$ are fitted to the three point-wise representations of \textit{ab-initio} PECs. This provides us with the three desired MLR functions representing the PECs from the three theoretical calculations, which we later use as starting conditions for fitting our experimental data. Since harmonic and first anharmonic contributions are sufficient to represent the data, see Sec.\,\ref{subsec:HotExpPart.Results}, only the parameters that strongly affect the lowest derivatives at $r=r_e$ need to be fitted to the thermoluminescence data. These are the coefficients $\beta_i$ with lowest $i$. In summary, in the following fitting procedure of all experimental data, we will treat $\beta_{0\leq i\leq 2}$, $C_6$, $C_8$ and $D_e$ as the only fitting parameters, retaining in this way the theoretical shape of each PEC in the region where no data are available.

A well defined representation of the $\mathrm{B}(2)^2\Sigma^+$ state is needed to simulate the thermoluminescence spectrum. We adopt a point-wise representation determined both by our experiment and by theoretical calculations in the region of missing data. This is realized by initializing the PEC with the \textit{ab-initio} predictions and adapting it to fitted Dunham coefficients via the Inverted Perturbation Approach \cite{IPAPASHOV}, see Sec.\,\ref{subsec:JointPart.Method}. The bottom part of the potential, determined by the experiment, and the upper part, determined by theory, are matched smoothly to provide a well depth referenced to that of the ground state. Within this representation the fitting parameters are the Dunham coefficients, which provide the link between the representations of the $\mathrm{X}(1)^2\Sigma^+$ and $\mathrm{B}(2)^2\Sigma^+$ states.

\subsection{Fit Method}
\label{subsec:JointPart.Method}
We fit our model of the $\mathrm{X}(1)^2\Sigma^+$ and $\mathrm{B}(2)^2\Sigma^+$ states to experimental data both from two-colour photoassociation and thermoluminescence/LIF spectroscopy. In particular, the fitted experimental quantities for two-colour PA are binding energies, while in the case of thermoluminescence/LIF they are band-head wavenumbers and, with lesser precision, the overall intensity profile. We recall that the fit parameters are those defining $V_{\mathrm{MLR}}$ for the $\mathrm{X}(1)^2\Sigma^+$ state and the Dunham coefficients of the $\mathrm{X}(1)^2\Sigma^+ - \mathrm{B}(2)^2\Sigma^+$ system. While the weakly-bound spectrum and the band-head positions do not determine precisely the equilibrium distances, the intensity profile carries this information together with the overall potential shapes and can be used to adjust the equilibrium distance of the $\mathrm{B}(2)^2\Sigma^+$ state with respect to that of the $\mathrm{X}(1)^2\Sigma^+$ state. The initial values for the fit parameters in $V_{\mathrm{MLR}}$ are defined in Sec.\,\ref{subsec:JointPart.Xstate} for each \textit{ab-initio} model, while the initial values for the Dunham coefficients are those of Table\,\ref{tab:RbSrDunham}. The figure of merit used in the fit is $\widetilde{\chi}^2$, see eqn\,(\ref{Chi-square}). In the following, we outline a single iteration step of our fit, which is applied to each \textit{ab-initio} model, while a future work will provide a detailed explanation \cite{AKAKE-I}.

We first generate the rovibrational levels of $\mathrm{X}(1)^2\Sigma^+$, using the fitted Dunham coefficients, for the range $v''= 0 - 6$ for $N''=0$, and fit them together with the experimental weakly-bound energy levels, via a direct potential fit of our model MLR PEC \cite{AKAKE-I}. We derive the $\mathrm{B}(2)^2\Sigma^+$ state depth from the MLR $D_e$ parameter and the Dunham coefficients. We then construct the $\mathrm{B}(2)^2\Sigma^+$ PEC via the Inverted Pertubation Approach, using both the $\mathrm{B}(2)^2\Sigma^+$ energy levels, generated with Dunham coefficients in the range $v'= 0 - 8$ for $N'= 0 - 44$, and the $\mathrm{B}(2)^2\Sigma^+$ potential well depth. We simulate the thermoluminescence spectra using the resulting PECs, in order to check the agreement of the simulated band-head positions and intensity profiles with the experimental ones. Here the convergence of the fit algorithm is checked and, if met, the calculation is stopped. Otherwise, we optimize the equilibrium point of the $\mathrm{B}(2)^2\Sigma^+$ state to maximize the agreement between the simulated intensity profile and the experimental one. During this optimization, for each change of the equilibrium point, the $\mathrm{B}(2)^2\Sigma^+$ state is optimized against the $\mathrm{X}(1)^2\Sigma^+$ state, which consists in fitting the Dunham coefficients of the $\mathrm{B}(2)^2\Sigma^+$ state keeping those of the $\mathrm{X}(1)^2\Sigma^+$ state fixed. With this new guess for the equilibrium distance of the $\mathrm{B}(2)^2\Sigma^+$ state, we refit all Dunham coefficients of both states, see Sec.\,\ref{subsec:HotExpPart.Results}, and repeat the iteration step.

\subsection{Results and discussion}
\label{subsec:JointPart.Results}
\begin{figure}[ht]
\centering
  \includegraphics[width=\columnwidth]{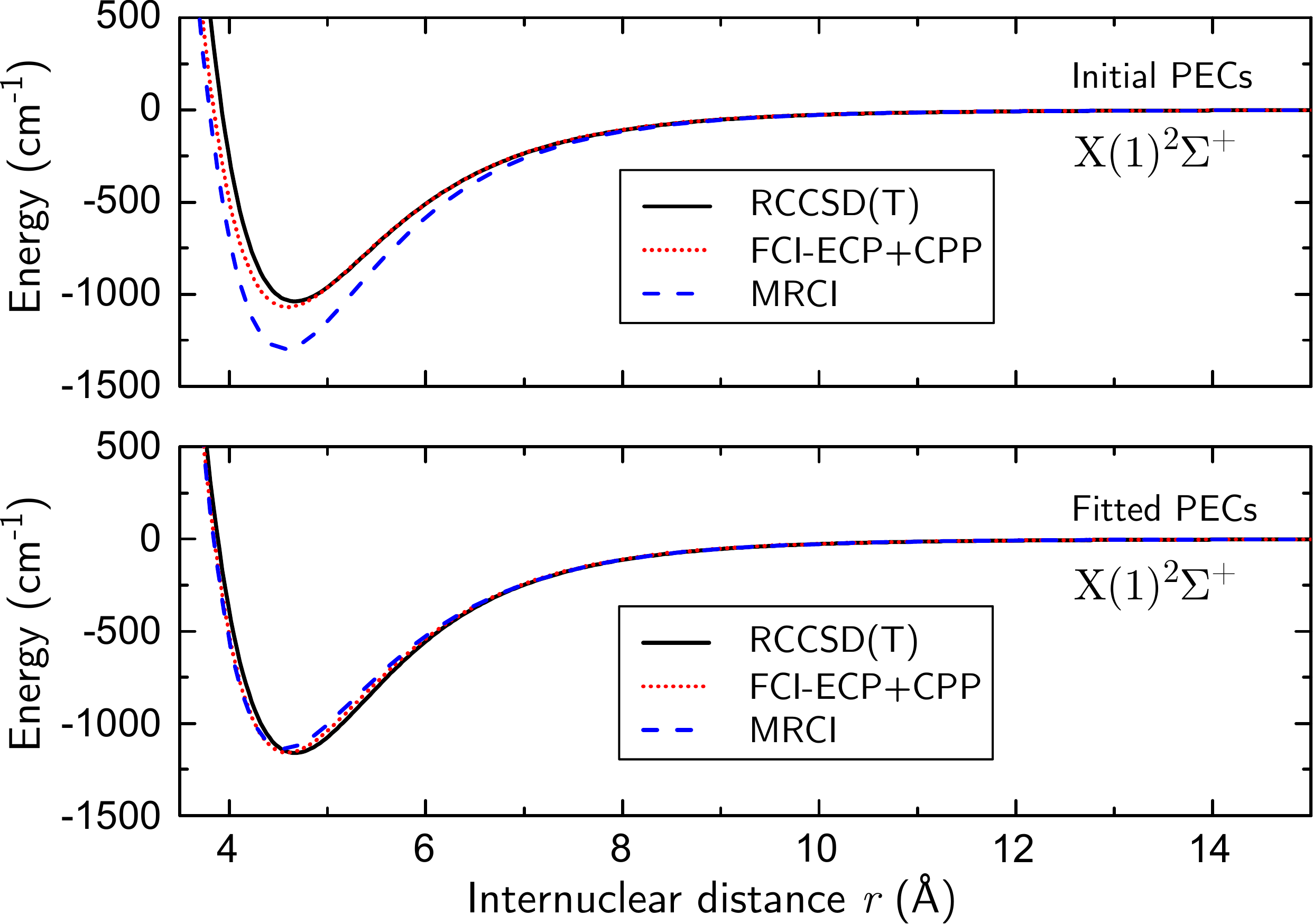}
  \caption{PECs of the $\mathrm{X}(1)^2\Sigma^+$ state of RbSr. Top panel: theoretical PECs corresponding to the three different \textit{ab-initio} calculations considered in this work. Bottom panel: PECs fitted to experimental data with initial fit parameters determined by each \textit{ab-initio} method, see Sec.\,\ref{subsec:JointPart.Method}.}
\label{fgr:MLR_BeforeAfterFIT}
\end{figure}

The fit outlined above is performed separately starting with FCI-ECP+CPP, RCCSD(T) and MRCI \textit{ab-initio} point-wise representations. In all cases we obtain good agreement between our best-fit model and the binding energies and band-head positions. However, while in the case of MRCI and FCI-ECP+CPP potential energy curves, the $\mathrm{B}(2)^2\Sigma^+$ state depth inferred after the first iteration is within $190\,\mathrm{cm}^{-1}$ of the \textit{ab-initio} predictions, in the case of RCCSDS(T) the well depth is about $440\,\mathrm{cm}^{-1}$ away from the theoretical value. As a consequence, we observe that all \textit{ab-initio} PECs give a sufficiently good representation of the RbSr ground state allowing for experimental fits, but only FCI-ECP+CPP and MRCI predictions are able to approximate the excited state well enough to permit its refinement by tuning its equilibrium distance. Best-fit parameters for the $\mathrm{X}(1)^2\Sigma^+$ state MLR functions and refined point-wise representations of the $\mathrm{B}(2)^2\Sigma^+$ state are reported in the Appendix \ref{sec:appendixPEC}. The derived Dunham coefficients for both $\mathrm{X}(1)^2\Sigma^+$ and $\mathrm{B}(2)^2\Sigma^+$ states are consistent with those in Table\,\ref{tab:RbSrDunham}. A comparison between the initial MLR functions, fitted to \textit{ab-initio} data, and the final MLR functions, based on \textit{ab-initio} PECs and fitted to experimental data, is shown in Fig.\,\ref{fgr:MLR_BeforeAfterFIT}.

The convergence of the three PECs towards a unique solution, as illustrated in Fig.\,\ref{fgr:MLR_BeforeAfterFIT}, and the good agreement with our experimental data corroborate our fit method. In particular, we observe that our data are sufficient to constrain strongly the depth of the ground state potential well to $D_e=1152_{-16}^{+9}$\,cm$^{-1}$ \footnote[1]{Although the depths derived from the FCI-ECP+CPP and RCCSD(T) methods are extremely close, we attribute to $D_e$ the mean value of all three cases and the full uncertainty range.}. The fitted PECs are consistent with the model-independent quantities derived in the previous sections up to residual model dependency. In particular, compared to those of Sec.\,\ref{sec:inclusionFFR}, the dispersion coefficient $C_6$ and the semiclassical phase $\Phi^0_{WKB}$ are slightly bigger, which is mostly due to the inclusion of $C_{10}$ in the MLR model \footnote[2]{By fitting once more the weakly-bound spectrum with $V_{LJ}$ including a $C_{10}$ term fixed to the theoretical value, we obtain $C_6=1.7962010665716115\times 10^7\,$\r{A}$^6\,\mathrm{cm}^{-1}$, $C_8=5.792504377056786\times 10^8\,$\r{A}$^8\,\mathrm{cm}^{-1}$ and $\Phi^0_{WKB}=67.4386$.}, while the $C_8$ is consistent within our relatively low precision. The fit quality of weakly-bound levels can still be assessed by the $\widetilde{\chi}^2$ and $\mathrm{DOF}=8$ used in Sec.\,\ref{sec:PrecisionFitPA}, since it is insensitive to $\beta_i$ fitting parameters. We obtain 0.89, 0.53 and 0.99 for MRCI, FCI-ECP+CPP and RCCSD(T), respectively, which are all sufficiently good. We use the $\widetilde{\chi}^2$, with $\mathrm{DOF}=10$ and error set to $0.16\;\mathrm{cm}^{-1}$, of the distance between band-head positions in experimental and simulated thermoluminescence spectra as a second  benchmark of the fitted potential energy curves. We obtain $0.83$, $0.5$ and $1.78$ for MRCI, FCI-ECP+CPP and RCCSD(T)  respectively, which shows agreement within our experimental resolution. 

In Table \ref{tab:RbSrConstantsX} the experimental values of spectroscopic constants are compared with the theoretical ones. Also here the convergence of the described fitting procedure is remarkable. The final value of the vibrational constant $\omega_{e}$ does not depend on the starting \textit{ab-initio} PECs used in the fit for both $\mathrm{X}(1)^2\Sigma^+$ and $\mathrm{B}(2)^2\Sigma^+$ states. However, the agreement between the experimental and theoretical $\omega_{e}$ values obtained is clearly the best for the FCI-ECP+CPP model. Similarly this model provides the best prediction of the potential well depths $D_{e}$ of the investigated states and of the number of bound levels in the ground state. The experimental data also allow to determine the difference between equilibrium distances of the $\mathrm{X}(1)^2\Sigma^+$ and $\mathrm{B}(2)^2\Sigma^+$ electronic states, $\Delta r_{e} = r_{e\mathrm{B}}-r_{e\mathrm{X}}$, and the values obtained are almost identical when starting from theoretical PECs calculated with the FCI-ECP+CPP ($\Delta r_{e} = -0.086$\AA) and MRCI ($\Delta r_{e} = -0.087$\AA) methods.

\begin{table*}[!ht]
	\caption{Comparison of spectroscopic constants and dispersion coefficients for the $\mathrm{X}(1)^{2}\Sigma^{+}$ and $\mathrm{B}(2)^{2}\Sigma^{+}$ states of the $^{85}\mathrm{Rb}^{88}\mathrm{Sr}$ molecule, between the present experiment-based work, the \textit{ab-initio} calculations used here \cite{ErnstLatestPECs, PiotrOlivierRbSr}, and other relevant works labelled as in their respective publication. Units of energy and length are $\mathrm{cm}^{-1}$ and \r{A}, respectively, while $\Phi^0_{WKB}$ is dimensionless. The final errors, defined as three times the standard deviation, are given in parentheses} 
	\label{tab:RbSrConstantsX}
\begin{tabular*}{\textwidth}{@{\extracolsep{\fill}}lllllllll}			
        \hline 
& $D_{e}$ & $\omega_{e}$ & $B_{e}\times10^2$ & $r_{e}$ [\AA{}]&$T_{e}$&$C_6\times10^{-7}$&$C_8\times 10^{-8}$&$\Phi^0_{WKB}$\\
\hline
        \multicolumn{8}{c}{$\mathrm{X}(1)^{2}\Sigma^{+}$}\\
       present$^{a}$&$1136$ & $40.39(72)$ & $1.874^{e}$ & $4.565^{e}$&$0$&$1.81(2)$&$5.8(1.3)$&$67.4393$\\            
       present$^b$ &$1158$ & $40.32(76)$ & $1.848^{e}$ & $4.595^{e}$&$0$&$1.80(2)$&$6.1(1.3)$&$67.4381$\\ 
       present$^c$ &$1161$ & $40.31(76)$ & $1.791^{e}$ &$ 4.669^{e}$&$0$&$1.81(2)$&$5.90(1.3)$&$67.4396$\\
       experimental \cite{ErnstHeNPRL}& $-$ & $42(5)$ & $-$ & $-$&$-$&$-$&$-$&$-$\\              
       MRCI \cite{ErnstLatestPECs}  & $1298$ & $42.5$ & $1.874$ & $4.565$&$0$&$-$&$-$&$70.7768$\\ 
       FCI-ECP+CPP \cite{PiotrOlivierRbSr}  & $1073.3$ & $38.98$ & $1.848$ & $4.595$&$0$&$-$&$-$&$65.8890$  \\
       RCCSD(T) \cite{PiotrOlivierRbSr} & $1040.5$& $38.09$ & $1.791$ & $4.669$&$0$ &$-$&$-$&$64.7252$\\
       ST \cite{Ernst2014JCP}& $1273$ & $42.2$ & $1.853$ & $4.590$&$0$&$-$&$-$&$-$\\
       CCSD(T) \cite{KajitaRbSr}& $916$ & $36$ & $1.75$ & $4.72$&$0$&$-$&$-$&$-$\\
      Relativistic KR-MRCI \cite{WangRbSr}& $1017.58$ & $35.8$ & $1.8$ & $4.66$&$0$&$-$&$-$&$-$\\
       theory \cite{CICPDispersionCoeffMitroy}&$-$ & $-$&$-$&$-$&$-$&$1.783$&$6.220$&$-$\\
     \hline
     \multicolumn{8}{c}{$\mathrm{B}(2)^{2}\Sigma^{+}$}\\
       present$^a$&$5025$ & $58.92(38)$ &$1.946$ &$4.478$ & $8848.0(8)$&$-$&$-$&$-$\\ 
       present$^b$ &$5047$ & $58.94(39)$ &$1.920 $& $4.509$ &$8847.6(8)$&$-$&$-$&$-$\\       
       present$^d$ &$5050$ & $58.95(39)$ & $1.893$ & $-$&$8847.7(8)$&$-$&$-$&$-$\\
       MRCI \cite{ErnstLatestPECs}  & $5214$ & $59.5$ & $1.921$ & $4.507$& $8830$&$-$&$-$&$-$\\    
       FCI-ECP+CPP \cite{PiotrOlivierRbSr}  & $4982.9$ & $58.37$ & $1.975$ & $4.445$&$8828$&$-$&$-$&$-$  \\
	   EOM-CC \cite{PiotrOlivierRbSr} & $4609.6$& $60.20$ & $1.925$ & $4.503$&$9224$&$-$&$-$&$-$ \\      
       ST \cite{Ernst2014JCP}& $5078$ & $58.5$ & $1.899$ & $4.533$& $8711$&$-$&$-$&$-$\\
     Relativistic KR-MRCI \cite{WangRbSr}& $4683.56$ & $58.1$ & $1.98$ & $4.43$&$9151$&$-$&$-$&$-$\\
       theory \cite{CICPDispersionCoeffMitroy}&$-$ & $-$&$-$&$-$&$-$&$8.448$&$59.80$&$-$\\
    \hline           
	\end{tabular*} \\
    $^{a}$ Based on MRCI \cite{ErnstLatestPECs} \textit{ab-initio }  calculation.\\
	$^{b}$ Based on FCI-ECP+CPP \cite{PiotrOlivierRbSr} \textit{ab-initio } calculation.  \\
    $^{c}$ Based on RCCSD(T) \cite{PiotrOlivierRbSr} \textit{ab-initio }  calculation. \\    
    $^{d}$ Based on RCCSD(T) \cite{PiotrOlivierRbSr} and EOM-CC \cite{PiotrOlivierRbSr} \textit{ab-initio } calculation; parameters taken from Dunham coefficients listed in Table \ref{tab:RbSrDunham}. \\
    $^{e}$ Fixed during the fit at the corresponding theoretical value. \\ 
\end{table*}

\begin{figure}[ht]
\centering
  \includegraphics[width=\columnwidth]{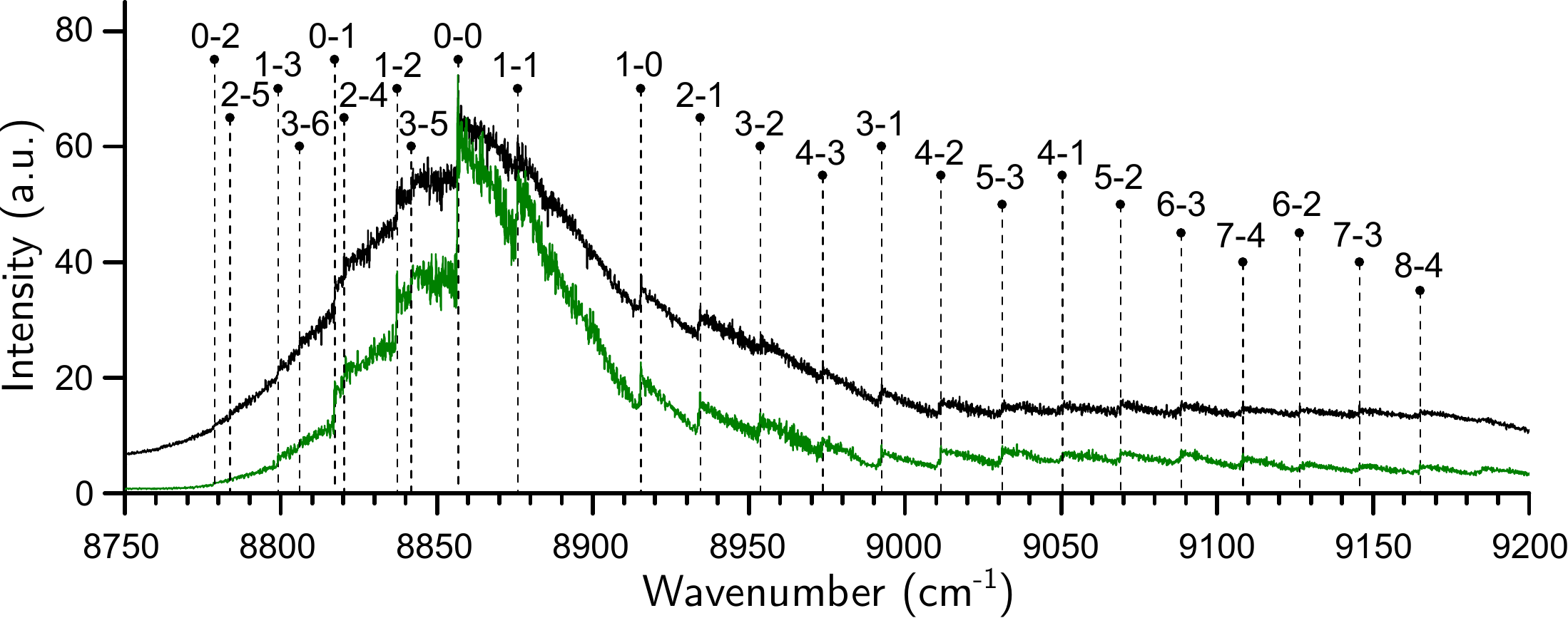}
  \caption{Comparison of the original experimental spectrum (upper curve, in black) with the spectrum simulated using the experimental potential employing the ground state $r_e$ value from the FCI-ECP+CPP \cite{PiotrOlivierRbSr} calculations (lower curve, in green). The positions of identified band heads are marked with dashed lines, on top of which the assigned vibrational quantum numbers $\nu'-\nu''$ are given.
}
  \label{fgr:RbSr_after_fit}
\end{figure}

Finally, we check in two ways the quality of the final fitted potential for the $\mathrm{X}(1)^2\Sigma^+$ state. Firstly, we simulate the thermoluminescence spectrum by using the potential we obtained starting from the FCI-ECP+CPP potential, as it gives the best agreement between theoretical and experimental values of molecular constants. In Fig.\,\ref{fgr:RbSr_after_fit} we show a comparison of this simulation with the experimental results. The agreement for the band-head positions between the two spectra is almost perfect, and this allows the assignment of even more band heads. Secondly, we use the fitted $\mathrm{X}(1)^2\Sigma^+$ state potential to calculate the positions of Fano-Feshbach resonances, which are listed in Table\,\ref{table:Feshbach Resonances}. At the time of the writing of this paper and thanks to these predictions, the resonances arising at about $1.3\,\mathrm{kG}$ for $^{87}\mathrm{Rb}^{84}\mathrm{Sr}$ and $1.0\,\mathrm{kG}$ for $^{87}\mathrm{Rb}^{88}\mathrm{Sr}$ have indeed been observed experimentally at the expected magnetic fields, which proves the high quality of the potential we obtained for the RbSr $\mathrm{X}(1)^2\Sigma^+$ state.

\begin{table*}[ht]
\tiny
\caption{Fano-Feshbach resonances for RbSr isotopologues due to coupling between $N=0$ molecular levels and $N=0$ atomic scattering levels in the magnetic field region $B < 3.0\,\mathrm{kG}$. $E_b$ is the binding energy of the bound state, $f^{Rb}$, $m_f^{Rb}$, $F$ and $m_F$ are the quantum numbers defined in the main text identifying the open and closed channel, respectively. $B$ is the resonant magnetic field, $\Delta \mu$ is the differential magnetic moment, and $\Delta B$ is the width of the resonance calculated as the avoided crossing gap divided by $\Delta \mu$. The avoided crossing gap is calculated for the two-body problem by using first order perturbation theory and the \textit{ab-initio} coupling matrix term induced by the variation of Rb hyperfine constant \cite{PiotrRbSrMagnetoassociation}. The two atoms are confined in a species-independent potential well with $\omega=2\pi \times 60\,\mathrm{kHz}$ isotropic trapping frequency, which is a typical value for sites of a three-dimensional optical lattice}
\renewcommand{\arraystretch}{0.065}
\begin{tabular*}{\textwidth}{@{\extracolsep{\fill}}llllllll}
\hline
BE\,(MHz) & $f^{Rb}$ & $m_f^{Rb}$ & $F$ & $m_F$ & $B\, \mathrm{(G)}$ & $\Delta \mu \,\mathrm{(MHz/G)}$ & $\Delta B\,\mathrm{(mG)}$\\ \hline
$^{85}\mathrm{Rb}^{84}\mathrm{Sr}$ &  &  &  &  &  &  & \\ \hline
6644.83 & 2 & -2 & 3 & -2 & 2950.99 & -2.64 & 31.40\\
6644.83 & 2 & -1 & 3 & -1 & 2500.46 & -2.53 & 35.06\\
6644.83 & 2 & 0 & 3 & 0 & 2108.76 & -2.49 & 31.82\\
6644.83 & 2 & 1 & 3 & 1 & 1778.42 & -2.53 & 24.94\\
6644.83 & 2 & 2 & 3 & 2 & 1506.91 & -2.64 & 16.03\\ \hline
$^{85}\mathrm{Rb}^{86}\mathrm{Sr}$ &  &  &  &  &  &  & \\ \hline
3421.32 & 2 & -2 & 3 & -2 & 1637.55 & -2.10 & 5.37\\
3421.32 & 2 & -1 & 3 & -1 & 1029.72 & -1.54 & 5.85\\
3421.32 & 2 & 0 & 3 & 0 & 562.87 & -1.29 & 4.03\\
3421.32 & 2 & 1 & 3 & 1 & 307.68 & -1.54 & 1.75\\
3421.32 & 2 & 2 & 3 & 2 & 193.47 & -2.10 & 0.63\\
9308.75 & 2 & 1 & 3 & 1 & 2799.16 & -2.67 & 4.71\\
9308.75 & 2 & 2 & 3 & 2 & 2499.41 & -2.72 & 3.26\\ \hline
$^{85}\mathrm{Rb}^{87}\mathrm{Sr}$ &  &  &  &  &  &  & \\ \hline
78.63 & 2 & -1 & 2 & -2 & 157.16 & -0.53 & 0\\
78.63 & 2 & 0 & 2 & -1 & 165.59 & -0.48 & 0\\
78.63 & 2 & 1 & 2 & 0 & 174.48 & -0.43 & 0\\
78.63 & 2 & 2 & 2 & 1 & 183.83 & -0.39 & 0\\
78.63 & 3 & -3 & 3 & -2 & 149.40 & -0.59 & 0\\
78.63 & 3 & -2 & 3 & -1 & 157.41 & -0.53 & 0\\
78.63 & 3 & -1 & 3 & 0 & 165.88 & -0.47 & 0\\
78.63 & 3 & 0 & 3 & 1 & 174.82 & -0.43 & 0\\
78.63 & 3 & 1 & 3 & 2 & 184.23 & -0.39 & 0\\
78.63 & 3 & 2 & 3 & 3 & 194.10 & -0.35 & 0\\
1071.24 & 2 & -2 & 3 & -3 & 1329.33 & 0.56 & 0\\
1071.24 & 3 & -3 & 3 & -2 & 995.37 & -1.85 & 0\\
4227.19 & 2 & -2 & 3 & -2 & 1995.89 & -2.37 & -5.26\\
4227.19 & 2 & -2 & 3 & -1 & 1741.03 & -2.22 & 0\\
4227.19 & 2 & -1 & 3 & -2 & 1740.39 & -2.23 & 0\\
4227.19 & 2 & -1 & 3 & -1 & 1470.84 & -2.06 & 5.63\\
4227.19 & 2 & -1 & 3 & 0 & 1256.20 & -1.98 & 0\\
4227.19 & 2 & 0 & 3 & -1 & 1255.68 & -1.99 & 0\\
4227.19 & 2 & 0 & 3 & 0 & 1049.46 & -1.95 & 4.51\\
4227.19 & 2 & 0 & 3 & 1 & 889.86 & -1.97 & 0\\
4227.19 & 2 & 1 & 3 & 0 & 889.49 & -1.98 & 0\\
4227.19 & 2 & 1 & 3 & 1 & 748.80 & -2.06 & 2.87\\
4227.19 & 2 & 1 & 3 & 2 & 642.03 & -2.19 & 0\\
4227.19 & 2 & 2 & 3 & 1 & 641.79 & -2.19 & 0\\
4227.19 & 2 & 2 & 3 & 2 & 551.81 & -2.37 & -1.46\\
4227.19 & 2 & 2 & 3 & 3 & 482.92 & -2.57 & 0\\
4227.19 & 3 & -3 & 3 & -2 & 2314.53 & -2.65 & 0\\
10827.35 & 2 & 2 & 3 & 3 & 2916.48 & -2.77 & 0\\ \hline
$^{85}\mathrm{Rb}^{88}\mathrm{Sr}$ &  &  &  &  &  &  & \\ \hline
5128.78 & 2 & -2 & 3 & -2 & 2364.10 & -2.52 & 8.32\\
5128.78 & 2 & -1 & 3 & -1 & 1879.36 & -2.33 & 9.05\\
5128.78 & 2 & 0 & 3 & 0 & 1474.80 & -2.26 & 7.75\\
5128.78 & 2 & 1 & 3 & 1 & 1157.32 & -2.33 & 5.57\\
5128.78 & 2 & 2 & 3 & 2 & 920.02 & -2.52 & -3.24\\ \hline
$^{87}\mathrm{Rb}^{84}\mathrm{Sr}$ &  &  &  &  &  &  & \\ \hline
9242.30 & 1 & 0 & 2 & 0 & 2218.82 & -1.89 & 3.77\\
9242.30 & 1 & 1 & 2 & 1 & 1312.74 & -2.15 & 1.69\\ \hline
$^{87}\mathrm{Rb}^{86}\mathrm{Sr}$ &  &  &  &  &  &  & \\ \hline
12546.80 & 1 & 1 & 2 & 1 & 2726.72 & -2.47 & 4.35\\ \hline
$^{87}\mathrm{Rb}^{87}\mathrm{Sr}$ &  &  &  &  &  &  & \\ \hline
0.01 & 1 & -1 & 1 & 0 & 0.18 & 0.70 & 0\\
0.01 & 1 & 0 & 1 & 1 & 0.18 & 0.70 & 0\\
0.01 & 2 & -1 & 2 & -2 & 0.18 & 0.70 & 0\\
0.01 & 2 & 0 & 2 & -1 & 0.18 & 0.70 & 0\\
0.01 & 2 & 1 & 2 & 0 & 0.18 & 0.70 & 0\\
0.01 & 2 & 2 & 2 & 1 & 0.18 & 0.70 & 0\\
287.32 & 1 & 0 & 1 & -1 & 397.14 & -0.74 & 0\\
287.32 & 1 & 1 & 1 & 0 & 432.35 & -0.62 & 0\\
287.32 & 2 & -2 & 2 & -1 & 366.06 & -0.88 & 0\\
287.32 & 2 & -1 & 2 & 0 & 398.65 & -0.73 & 0\\
287.32 & 2 & 0 & 2 & 1 & 434.30 & -0.62 & 0\\
287.32 & 2 & 1 & 2 & 2 & 473.02 & -0.52 & 0\\
1952.06 & 2 & -2 & 2 & -1 & 1672.34 & -1.70 & 0\\
6235.49 & 1 & -1 & 2 & -2 & 294.64 & 1.96 & 0\\
6235.49 & 1 & -1 & 2 & -1 & 1918.11 & -0.88 & -59.55\\
6235.49 & 1 & -1 & 2 & -1 & 519.45 & 0.88 & 16.13\\ \hline
$^{87}\mathrm{Rb}^{88}\mathrm{Sr}$ &  &  &  &  &  &  & \\ \hline
7403.10 & 1 & -1 & 2 & -1 & 2804.54 & -1.68 & -8.73\\
7403.10 & 1 & 0 & 2 & 0 & 1014.50 & -1.08 & -5.70\\
7403.10 & 1 & 1 & 2 & 1 & 366.98 & -1.68 & -1.14\\ \hline\label{table:Feshbach Resonances}
\end{tabular*}
\end{table*}

\section{Conclusions and Outlook}
\label{sec:Conclusions}

We have performed three different types of spectroscopy experiments in order to investigate the ground and second excited electronic states, both of $^{2}\Sigma^{+}$ symmetry, of the alkali -- alkaline-earth RbSr molecule. We have presented a novel procedure to connect data from two-colour photoassociation measurements, which provide information on energy levels of the $\mathrm{X}(1)^{2}\Sigma^{+}$ state near the dissociation threshold, and low-resolution data from thermoluminescence/LIF experiments, which allow to describe the bottom of both $\mathrm{X}(1)^{2}\Sigma^{+}$ and $\mathrm{B}(2)^{2}\Sigma^{+}$ PECs. As the thermoluminescence spectra lack rotational resolution, the equilibrium distance between the Rb and Sr nuclei cannot be determined from our measurements and must be taken from theoretical calculations. Therefore we use three different sets of theoretical PECs, resulting from state-of-the-art \textit{ab-initio} calculations, as starting points for the fit of potential energy curves to the experimental data. We obtain three potentials for the $\mathrm{X}(1)^{2}\Sigma^{+}$ state, but despite significant differences between the starting potentials, the three fitted ones converge to nearly the same shape. In the region of missing experimental data the shapes of the fitted PECs stay close to the initial theoretical potentials and this region awaits future spectroscopic investigation to be refined. Although in the case of the $\mathrm{B}(2)^{2}\Sigma^{+}$ state, the experimental data provide only information about deeply-bound energy levels, our procedure is able to reject one of the three theories that diverges too much from the experimental results.

We have demonstrated that our data analysis method is a powerful tool to obtain potential energy curves of heavy molecules, where achieving rotational resolution is difficult and investigation of the mid-range spectrum challenging. This method may find a welcome use in the field of physical chemistry, since it shows, in the simple case of diatomic molecules, that several independent sources of information, both experimental and theoretical, can be synthesized successfully. The findings of our analysis may be of interest to physicists from various fields. Indeed, calculations performed with the fitted RbSr potentials demonstrate their power to predict the positions of unassigned band heads and the intensity distribution of the spectrum, but also the positions of Fano-Feshbach resonances, some of which were later confirmed experimentally \cite{RbSrFFR}. As a next step, we plan to further refine the PECs derived in this work via new LIF experiments with rotational resolution, and to characterize the effects induced by hyperfine and spin-rotation couplings via additional two-color PA. We will also use the results of this work to determine an efficient STIRAP path for RbSr molecules towards the rovibronic ground state \cite{WangRbSr}. With such molecules available, one can run quantum simulations \cite{Perez-Rios2010MagneticExciton, Micheli2006ToolboxPolMol, Baranov2012RevDipQGases,Bohn1002RevColdMol}, perform fundamental tests of physics \cite{Cahn2014ParityViolation, Kajita2008ProtonTonElectronMass2Sigma, Safronova2017RevNewPhysicsAtMol}, and study chemical reactions with full control over reactants at the quantum level \cite{Wolf2017StateToStateChemistry, Ospelkaus2010ChemReactKRb, Sikorsky2018SpinAtomIonChem, Krems2008RevColdChemistry, Bohn1002RevColdMol}.

\section{Acknowledgements}
This project has received funding from the European Research Council (ERC) under the European Union's Seventh Framework Programme (FP7/2007-2013) (Grant agreement No. 615117 QuantStro). B.P. thanks the NWO for funding through Veni Grant No. 680-47-438 and C.-C.C. thanks the Ministry of Education of the Republic of China (Taiwan) for a MOE Technologies Incubation Scholarship. J.S. acknowledges partial support from the Miniatura I  programme founded by the National Science Centre of Poland (Grant no. 2017/01/X/ST2/00057). This work was partially supported by the National Science Centre of Poland (Grant no. 2016/21/B/ST2/02190).  A.C. is grateful to Jeremy Hutson for fruitful discussions. J.S. is grateful to Olivier Dulieu and Piotr \.{Z}uchowski for fruitful discussions.

\clearpage
\section{Appendix}

\subsection{Theoretical model for inter-species thermalization}
\label{sec:CrossThermalizationTheory}
A rigorous analysis of the experimental data would require a Monte-Carlo trajectory simulation taking into account the initial atomic distributions, the subsequent excitation of the Rb cloud and the elastic-scattering cross sections, both inter-species and intra-species. However, we are here only interested in a confirmation of our ground-state potential model and for that we do not require precise values for the scattering lengths. Moreover we observe that, although during the thermalization the system is out of equilibrium, it might be close enough to equilibrium to apply a very simple collision model giving the cross-thermalization rate from equilibrium statistical physics \cite{MoskCrossThermalization}. This is suggested in our case by the very fast decrease in Rb temperature compared to the cross-thermalization time. In order to clarify the analysis, we review the model here.

The temperature difference is expected to decrease exponentially to zero with an inter-species thermalization rate given by

\begin{equation}
\label{CrossThermRate}
\tau^{-1}=\frac{d(\Delta T)}{\Delta T dt}=\frac{d(\Delta T)}{\Delta T}\Gamma
,
\end{equation}

\noindent
where $\Delta T$ is the temperature difference between species 1 and 2 and $dt=1/\Gamma$ is the average collision time. The rate of inter-species collisions $\Gamma$ is given by

\begin{equation*}
\Gamma=\sigma_{12}\times\bar{v}\times\int{n_1(x)\,n_2(x)\,dx^3}=
\end{equation*}
\begin{equation}
\label{InterspeciesCollisionRate}
=\sigma_{12}\times\bar{v}\times N_1 N_2 \int{\rho_1(x)\,\rho_2(x)\,dx^3},
\end{equation}

\noindent
where $\sigma_{12}=4\pi a^2_{12}$ is the inter-species cross section, $\bar{v}$ is the mean thermal relative velocity and $n_{1,2}(x)$, $\rho_{1,2}(x)$ are the atomic density distributions normalized to $N_{1,2}$ or $1$, respectively. At thermal equilibrium with known trapping potential $U_{1,2}$, temperatures $T_{1,2}$, and atomic masses $m_{1,2}$, we know all the quantities in the equation above except the inter-species scattering length. In particular $\bar{v}=\sqrt{(8 k_B/\pi) \times((T_1/m_1)+(T_2/m_2))}$ and $n_{1,2}(x)\propto \exp^{-U_{1,2}(x)/k_B T_{1,2}}$. 

From basic kinematics the energy transfer from species 1 to species 2 is  given by

\begin{equation*}
\Delta E_{1 \rightarrow 2} = \xi \, k_B \Delta T,
\end{equation*}
\begin{equation}
\label{EnergyTransfer}
\xi=\frac{4 m_1 m_2}{(m_1 + m_2)^2},
\end{equation}

\noindent
where $\xi$ accounts for the mass imbalance, and in our case $\xi\simeq 1$. From this we obtain

\begin{equation}
\label{TemperatureChange}
d(\Delta T)= \frac{\xi}{3} \frac{N_1+N_2}{N_1 N_2} \Delta T.
\end{equation}

Substituting eqn\,(\ref{TemperatureChange}) into eqn\,(\ref{CrossThermRate}), we get the final result

\begin{equation*}
\tau^{-1}= \frac{\xi}{3} \frac{N_1+N_2}{N_1 N_2} \Gamma = \frac{\xi}{3} \sigma_{12}\times (N_1+N_2)\,\bar{v}  \,\int{\rho_1(x)\,\rho_2(x)\,dx^3}=
\end{equation*}

\begin{equation}
\label{CrossThermRate2}
=\frac{\xi}{3} \sigma_{12}\times \Phi,
\end{equation}

\noindent
where the kinematic contribution to the rate is summarized in the effective flux $\Phi$.

The value of 3 in the denominator of eqn\,(\ref{CrossThermRate2}) represents the average number of collisions for thermalization. Corrections to this number have been evaluated \cite{DalibardColdCollisions}, and it is shown to vary within the range $2.4-3.4$, with $2.4$ referring to fast thermalization compared to trap oscillation time and $3.4$ to the opposite case.

\subsection{Potential energy curves}
\label{sec:appendixPEC}
\begin{figure}[ht]
\centering
  \includegraphics[width= \columnwidth]{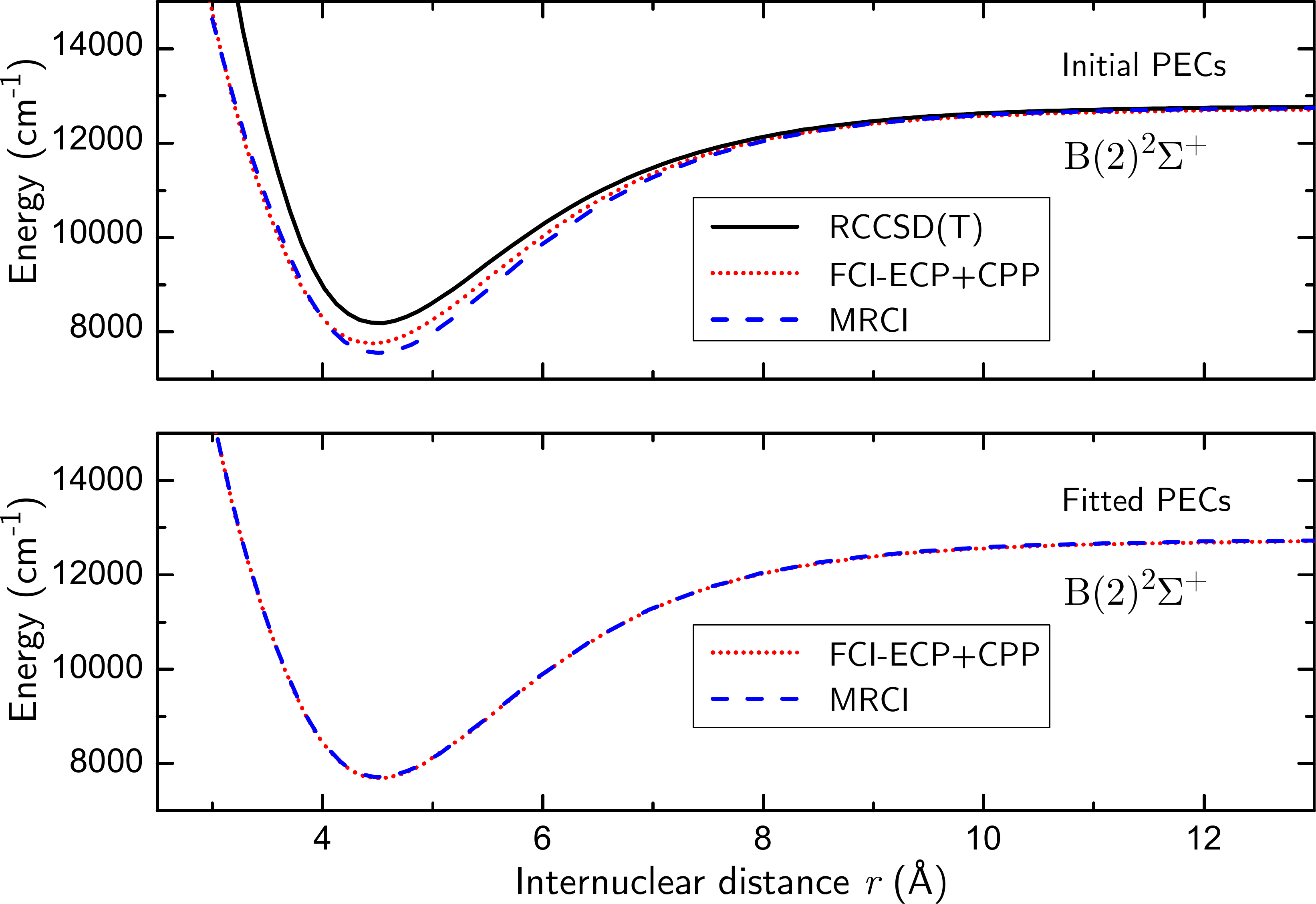}
  \caption{PECs of the $\mathrm{B}(2)^2\Sigma^+$ state of RbSr. Top panel: theoretical PECs corresponding to the three different \textit{ab-initio} calculations considered in this work. Bottom panel: PECs fitted to experimental data with initial fit parameters determined by the FCI-ECP+CPP and MRCI methods, see Sec.\,\ref{subsec:JointPart.Method}.}
\label{fgr:B_BeforeAfterFIT} 
\end{figure}

In this part, we provide additional information about the fitted potentials. In Table\,\ref{tab:BestFitParameters_MLR}, we give the best fit parameters for the MLR PEC describing the $\mathrm{X}(1)^2\Sigma^+$ state. In Table\,\ref{tab:PWforBstateFCI} and Table\,\ref{tab:PWforBstateMRCI} we give the point-wise representations of the fitted PECs for the $\mathrm{B}(2)^2\Sigma^+$ state, fitted starting from the FCI-ECP+CPP and the MRCI methods, respectively \footnote[1]{Electronic Supplementary Information (ESI) available. See \url{http://dimer.ifpan.edu.pl/}.}. Finally, in Fig.\,\ref{fgr:B_BeforeAfterFIT} we show a comparison of the potentials for the $\mathrm{B}(2)^2\Sigma^+$ state before and after our fit procedure.

\begin{table*}[ht]
	\caption{Best fit parameters for the MLR PECs describing the $\mathrm{X}(1)^2\Sigma^+$ state in the cases of the three initial \textit{ab-initio} representations. Units of energy and length are cm$^{-1}$ and \r{A}, respectively. The number of digits in the presented values of parameters is necessary to reproduce band-head positions and weakly-bound energy levels with the experimental uncertainty}
	\label{tab:BestFitParameters_MLR}
	\begin{tabular*}{\textwidth}{@{\extracolsep{\fill}}llll}	
		\hline
        \noalign{\smallskip}
	 & MRCI \cite{ErnstLatestPECs} & FCI-ECP+CPP \cite{PiotrOlivierRbSr}& RCCSD(T) \cite{PiotrOlivierRbSr}\\ \hline
        \noalign{\smallskip}
		\multicolumn{1}{c}{$p$} & $5^a$ & $5^a$ & $5^a$\\
        \multicolumn{1}{c}{$q$} & $2^a$ & $2^a$ & $2^a$\\
        \multicolumn{1}{c}{$N_u$} & $3^a$ & $3^a$ & $3^a$\\
        \multicolumn{1}{c}{$N_{\beta}$} & $5^a$ & $5^a$ & $5^a$\\\hline
        \noalign{\smallskip}
        \multicolumn{1}{c}{$D_e$} & $1136.153957156767$ & $1158.334744879383$ & $1161.0696743991873$\\
        \multicolumn{1}{c}{$r_e$} & $4.5645^a$ & $4.59508^a$ & $4.66879^a$\\
		\multicolumn{1}{c}{$C_6$ } & $1.808868014576728$ & $1.795668695101867$ & $1.8134615231939677$ \\
		\multicolumn{1}{c}{$C_8$} & $5.792504377056786$ & $6.148308472469144$ & $5.870256113661574$ \\
        \multicolumn{1}{c}{$C_{10}$} & $2.2043858534998264^a$ & $2.2043858534998264^a$ & $2.2043858534998264^a$ \\
        \multicolumn{1}{c}{$\beta_0$} & $-1.2521217820591306$ & $-1.2744532761179883$ & $-1.2541965574219214$ \\
        \multicolumn{1}{c}{$\beta_1$} & $-2.7403962754860123$ & $-2.6486718159733136$ & $-2.324690124968812$ \\
        \multicolumn{1}{c}{$\beta_2$} & $-1.2388430923676004$ & $-0.8587136858852784$ & $-0.06921139967893859$ \\
        \multicolumn{1}{c}{$\beta_3$} & $0.8220377227516734^a$ & $1.5773120878976061^a$ & $0.9325813112428021^a$ \\
        \multicolumn{1}{c}{$\beta_4$} & $-2.710995726338915^a$ & $-0.16154919058041997^a$ & $-4.2183716802600495^a$ \\
        \multicolumn{1}{c}{$\beta_5$} & $-4.142301068756231^a$ & $-0.8834374478517256^a$ & $-5.290318777716273^a$ \\\hline
	\end{tabular*} \\
    $^{a}$ Held fixed during fit to experimental data.\\
\end{table*}

\begin{table}[ht]
\caption{The point-wise potential energy curve of the $B(2)^2\Sigma^+$ state obtained after the fit procedure, based on the theoretical curve calculated with the FCI-ECP+CPP method}
\label{tab:PWforBstateFCI}
\renewcommand{\arraystretch}{0.5}
\begin{tabular*}{\columnwidth}{@{\extracolsep{\fill}}llll}
\hline
$r \,($\AA{}$)$ & $E\,(\mathrm{cm}^{-1})$ & $r \,($\AA{}$)$ & $E\,(\mathrm{cm}^{-1})$ \\
\hline
  2.68 &  19257.0300 &   8.18 &  12109.8735 \\
  2.78 &  17908.5708 &   8.29 &  12157.4315 \\
  2.89 &  16617.8234 &   8.39 &  12197.4666 \\
  3.00 &  15379.9026 &   8.50 &  12238.2371 \\
  3.10 &  14312.2244 &   8.60 &  12272.5381 \\
  3.21 &  13241.3574 &   8.71 &  12307.4494 \\
  3.31 &  12375.1023 &   8.82 &  12339.6227 \\
  3.42 &  11528.6439 &   8.92 &  12366.6696 \\
  3.52 &  10837.7736 &   9.03 &  12394.1795 \\
  3.63 &  10150.2670 &   9.13 &  12417.2977 \\
  3.74 &   9536.4284 &   9.24 &  12440.8047 \\
  3.84 &   9050.4423 &   9.35 &  12462.4492 \\
  3.95 &   8604.8603 &   9.45 &  12480.6336 \\
  4.05 &   8289.8226 &   9.56 &  12499.1204 \\
  4.16 &   8021.3425 &   9.66 &  12514.6521 \\
  4.27 &   7838.3481 &   9.77 &  12530.4436 \\
  4.37 &   7737.9870 &   9.87 &  12543.7127 \\
  4.48 &   7691.2318 &   9.98 &  12557.2074 \\
  4.58 &   7699.9253 &  10.09 &  12569.6388 \\
  4.69 &   7753.0775 &  10.19 &  12580.0904 \\
  4.79 &   7837.0308 &  10.30 &  12590.7265 \\
  4.90 &   7961.2580 &  10.40 &  12599.6736 \\
  5.32 &   8634.7155 &  10.51 &  12608.7842 \\
  5.43 &   8837.5343 &  10.62 &  12617.1902 \\
  5.54 &   9044.9495 &  10.72 &  12624.3507 \\
  5.64 &   9233.5924 &  10.83 &  12631.8586 \\
  5.75 &   9439.0045 &  10.93 &  12638.2067 \\
  5.85 &   9621.9672 &  11.04 &  12644.7041 \\
  5.96 &   9818.0546 &  11.14 &  12650.2014 \\
  6.06 &   9991.0355 &  11.25 &  12655.8314 \\
  6.17 &  10174.3255 &  11.36 &  12661.0571 \\
  6.28 &  10349.5430 &  11.46 &  12665.4829 \\
  6.38 &  10501.3856 &  11.57 &  12670.0202 \\
  6.49 &  10659.9119 &  11.67 &  12673.8658 \\
  6.59 &  10796.1676 &  11.78 &  12677.8112 \\
  6.70 &  10937.3983 &  11.89 &  12681.4802 \\
  6.81 &  11069.6823 &  11.99 &  12684.5934 \\
  6.91 &  11182.3688 &  12.10 &  12687.7911 \\
  7.02 &  11298.2776 &  12.20 &  12690.5065 \\
  7.12 &  11396.6341 &  12.31 &  12693.2978 \\
  7.23 &  11497.4813 &  12.41 &  12695.6701 \\
  7.33 &  11582.8320 &  12.52 &  12698.1107 \\
  7.44 &  11670.1576 &  12.63 &  12700.3871 \\
  7.55 &  11751.0295 &  12.73 &  12702.3241 \\
  7.65 &  11819.3069 &  12.84 &  12704.3196 \\
  7.76 &  11889.0220 &  12.94 &  12706.0191 \\
  7.86 &  11947.8194 &  13.05 &  12707.7714 \\
  7.97 &  12007.8007 &  13.16 &  12709.4095 \\
  8.08 &  12063.2037 &        &             \\
  \hline
  \end{tabular*}
\end{table}

\begin{table}[ht]
\caption{The point-wise potential energy curve of the $B(2)^2\Sigma^+$ state obtained after the fit procedure, based on the theoretical curve calculated with the MRCI method}
\label{tab:PWforBstateMRCI}
\renewcommand{\arraystretch}{0.5}
\begin{tabular*}{\columnwidth}{@{\extracolsep{\fill}}llll}
\hline
$r \,($\AA{}$)$ & $E\,(\mathrm{cm}^{-1})$ & $r \,($\AA{}$)$ & $E\,(\mathrm{cm}^{-1})$ \\ \hline
  2.57 &  20015.9638 &   7.97 &  12039.5764 \\
  2.67 &  18623.1081 &   8.47 &  12260.0669 \\
  2.77 &  17378.0065 &   8.97 &  12412.7431 \\
  2.87 &  16227.5400 &   9.47 &  12513.9446 \\
  2.97 &  15164.7084 &   9.97 &  12581.4056 \\
  3.07 &  14191.0980 &  10.47 &  12626.6346 \\
  3.17 &  13301.5037 &  10.97 &  12657.3393 \\
  3.27 &  12484.9011 &  11.97 &  12693.3984 \\
  3.37 &  11729.7060 &  12.97 &  12711.7395 \\
  3.47 &  11028.5584 &  13.97 &  12721.6836 \\
  3.57 &  10381.0891 &  14.00 &  12721.9028 \\
  3.67 &   9791.4292 &  14.10 &  12722.6076 \\
  3.77 &   9271.8458 &  14.20 &  12723.2748 \\
  3.87 &   8816.5037 &  14.30 &  12723.9066 \\
  3.97 &   8486.8951 &  14.40 &  12724.5053 \\
  4.07 &   8191.1429 &  14.50 &  12725.0727 \\
  4.17 &   7968.0412 &  14.60 &  12725.6108 \\
  4.27 &   7821.7664 &  14.70 &  12726.1213 \\
  4.37 &   7739.7466 &  14.80 &  12726.6058 \\
  4.47 &   7711.9908 &  14.90 &  12727.0659 \\
  4.57 &   7729.6107 &  14.97 &  12727.3742 \\
  4.67 &   7784.9552 &  15.00 &  12727.5029 \\
  4.77 &   7871.4749 &  15.10 &  12727.9182 \\
  4.87 &   7983.6310 &  15.20 &  12728.3130 \\
  4.97 &   8116.4157 &  15.30 &  12728.6885 \\
  5.07 &   8265.5303 &  15.40 &  12729.0457 \\
  5.17 &   8430.3232 &  15.50 &  12729.3857 \\
  5.27 &   8609.5386 &  15.60 &  12729.7094 \\
  5.37 &   8792.9576 &  15.70 &  12730.0178 \\
  5.47 &   8981.4689 &  15.80 &  12730.3115 \\
  5.77 &   9548.1877 &  15.90 &  12730.5915 \\
  5.97 &   9908.3732 &  16.00 &  12730.8585 \\
  6.47 &  10690.4487 &  18.00 &  12734.2857 \\
  6.97 &  11287.8051 &  20.00 &  12735.7680 \\
  7.47 &  11725.8313 &        &             \\
  \hline
  \end{tabular*}
\end{table}

\clearpage
\bibliography{RbSrGroundStateBibliography}

\end{document}